\documentclass[aps,prd,onecolumn,nofootinbib,preprintnumbers,superscriptaddress,longbibliography,nobibnotes]{revtex4-1}
\usepackage{style}

\begin{document}

\title{Direct Deflection of Millicharged Radiation}

%%%%%%%%%%%%%%%%%%%%%%%%%%%%%%%%%%%%%%%%%%%%%%%%%%%%%%%%%%%%
\author{Asher Berlin}
\email{aberlin@fnal.gov}
\affiliation{Theoretical Physics Division, Fermi National Accelerator Laboratory, Batavia, IL 60510, USA}
\affiliation{Superconducting Quantum Materials and Systems Center (SQMS), Fermi National Accelerator Laboratory, Batavia, IL 60510, USA}
\author{Surjeet Rajendran}
\email{srajend4@jhu.edu}
\affiliation{The William H.~Miller III Department of Physics and Astronomy, The Johns Hopkins University, Baltimore, Maryland, 21218, USA}
\author{Harikrishnan Ramani}
\email{hramani@udel.edu}
\affiliation{Department of Physics and Astronomy, University of Delaware, Newark, DE 19716, USA}
\affiliation{Stanford Institute for Theoretical Physics, Stanford University, Stanford, California, 94305, USA}
\author{Erwin H.~Tanin}
\email{ehtanin@stanford.edu}
\affiliation{Stanford Institute for Theoretical Physics, Stanford University, Stanford, California, 94305, USA}
\affiliation{The William H.~Miller III Department of Physics and Astronomy, The Johns Hopkins University, Baltimore, Maryland, 21218, USA}
%%%%%%%%%%%%%%%%%%%%%%%%%%%%%%%%%%%%%%%%%%%%%%%%%%%%%%%%%%%%

\begin{abstract}
\vspace{0.3cm}
\noindent
Millicharged particles are generic in theories of dark sectors. 
A cosmic or local abundance of them may be produced by the early universe, stellar environments, or the decay or annihilation of dark matter/dark energy. Furthermore, if such particles are light, these production channels result in a background of millicharged radiation. We show that light-shining-through-wall experiments employing superconducting RF cavities can also be used as ``direct deflection" experiments to search for this relativistic background. The millicharged plasma is first subjected to an oscillating electromagnetic field of a driven cavity, which causes charge separation in the form of charge and current perturbations. In turn, these perturbations can propagate outwards and resonantly excite electromagnetic fields in a well-shielded cavity placed nearby, enabling detection. We estimate that future versions of the existing Dark SRF experiment can probe orders of magnitude of currently unexplored parameter space, including millicharges produced from the Sun, the cosmic neutrino background, or other mechanisms that generate a thermal abundance with energy density as small as $\sim 10^{-4}$ that of the cosmic microwave background. 
\end{abstract}

%%%%%%%%%%%%%%%%%%%%%%%%%%%%%%%%%%%%%%%%%%%%%%%%%%%%%%%%%%%%

\maketitle

{
\hypersetup{linkcolor=black}
\tableofcontents
}

\vspace{5mm}
\begin{center}
\emph{Conventions and Notation}
\end{center}

In this work, we use a mostly-negative spacetime metric $\eta^{\mu \nu} = \text{diag} (+1 , -1, -1, -1)$ and natural units, $\hbar = c = k_B = 1$, with rationalized Heaviside--Lorentz units for electromagnetic fields. We define the Fourier transform $f(k)$ of a function $f(x)$, as well as its inverse transform, as
\be
f(k) = \int d^4x ~ e^{i k \cdot x} \, f(x)
~~,~~ 
f(x) = \int \frac{d^4k}{(2\pi)^4} ~ e^{-i k \cdot x} \, f(k)
~.
\ee
For a four-momentum $k^\mu = (\w , \kv)$, we denote $k^2=\w^2-|\kv|^2$. Throughout, we will also adopt the notation where prime ``$\p$" superscripts are meant to indicate dark sector quantities. Furthermore, tildes, as in $\tilde{f} (k)$ and $\tilde{f} (x)$, indicate that a function (including its argument) is evaluated in the rest frame of the plasma, whereas the absence of a tilde corresponds to the laboratory frame.

%%%%%%%%%%%%%%%%%%%%%%%%%%%%%%%%%%%%%%%%%%%%%%%%%%%%%%%%%%%%
\newpage
\section{Introduction}
\label{sec:intro}
%%%%%%%%%%%%%%%%%%%%%%%%%%%%%%%%%%%%%%%%%%%%%%%%%%%%%%%%%%%%

Throughout its history, the universe has been an efficient factory for visible radiation. Most notably, we see this today in the form of the cosmic microwave background, starlight, and high-energy cosmic rays, which are efficiently produced by the hot and dense plasmas found in the early universe, stellar environments, or other late-time astrophysical systems. The efficiency of such reactions implies that even extremely feebly-coupled particles could be similarly produced. This is known to be the case for solar and supernova neutrinos, and expected to hold for neutrinos produced thermally in the early universe, constituting the cosmic neutrino background. 

It is thus generic to expect that new feebly-coupled particles beyond the Standard Model (SM) could be sourced in large numbers through analogous cosmological or astrophysical processes~\cite{Raffelt:1996wa,Bogorad:2021uew, Eby:2024mhd, Nguyen:2023czp,DeRocco:2019jti}. Such ``dark radiation" is typically searched for in one of two ways. First, for primordial radiation, its gravitational coupling can leave indirect telltale signatures in cosmological observations of, e.g., the cosmic microwave background~\cite{Hou:2011ec,Baumann:2015rya} and the abundance of light nuclei~\cite{Nollett:2013pwa,Nollett:2014lwa,Giovanetti:2024eff}. Second, if dark radiation possesses non-gravitational interactions with the SM, it can be detected directly with sensitive calorimetric or electromagnetic sensors. Although this latter strategy has been most actively explored in the development of helioscopes searching for new particles emitted by the Sun~\cite{Sikivie:1983ip,vanBibber:1988ge,Redondo:2008aa,Berlin:2021kcm,Irastorza:2011gs,CAST:2017uph} (and to some degree primordial dark radiation~\cite{Cui:2017ytb,Kuo:2021mtp,Dror:2021nyr,ADMX:2023rsk}), such a program has not been as widely pursued.

Millicharged particles (mCPs), i.e., particles with an effective electromagnetic charge $q_\x$ much smaller than that of the electron, are a generic class of dark sector particles. This is because it is reasonable that the dark sector has a massless (or approximately massless) $U(1)^\p$ gauge boson with stable matter that is charged under it. This $U(1)^\p$ can kinetically-mix with our photon, resulting in the dark sector matter acquiring a small effective electromagnetic charge~\cite{Holdom:1985ag}. If these matter particles are light, it is reasonable that there is a relativistic abundance of them today, manifesting as dark radiation. In this paper, we devise a strategy to detect such millicharged dark radiation. 

In particular, we focus on light-shining-through-wall (LSW) experiments. Although the original aim of such experiments is to directly produce and detect new light particles coupled to electromagnetism, we show that they are also inadvertently sensitive to a relativistic background of mCPs. Millicharged particles are deflected as they pass through a driven electromagnetic field, setting up collective phase-space disturbances that can propagate into a shielded region and excite small signal fields. Such a ``direct deflection" setup using quasistatic lumped element LC circuits was originally proposed in Ref.~\cite{Berlin:2019uco} to detect millicharged dark matter. Here, we show that LSW experiments using pairs of radio-frequency (RF) cavities can operate in a similar manner to search for millicharged radiation. We focus specifically on the sensitivity of superconducting RF (SRF) cavities, such as those used in the Dark SRF LSW experiment~\cite{Graham:2014sha,Romanenko:2023irv}, since these can achieve extremely large quality factors, $Q \sim \text{few} \times 10^{11}$~\cite{Romanenko:2014yaa}. Large $Q$-factors resonantly enhance the strength of the driven electromagnetic field and the detectable signal field, both of which benefit the overall sensitivity of a LSW experiment. While we focus solely on the prospects to detect a background of millicharged radiation (relativistic particles), we note that related works have previously explored the capability of Dark SRF to detect millicharged dark matter~\cite{Berlin:2023gvx} (non-relativistic particles), as well as directly produce and detect ultralight mCPs~\cite{Berlin:2020pey} (a controlled source of new particles, as opposed to an isotropic background).

The remainder of this work is as follows. In \Sec{overview}, we provide a conceptual overview of the class of models and signals discussed throughout this work. In \Sec{formalism}, we summarize the formalism used to determine the response of a millicharged plasma to a driven electromagnetic field similar to those used in RF cavities. Here, we also provide a general discussion on the sensitivity of a LSW experiment, and list the experimental assumptions that enter into our projections. The results of these detailed calculations are then applied to a couple of concrete examples of millicharged dark radiation. In \Sec{darksolarwind}, we consider millicharged radiation that is produced from the Sun and thermalizes through self-interactions. Then, in \Sec{cosmo}, we apply our formalism to millicharged radiation that arises cosmologically, such as from dark matter decay or annihilation, a dynamical dark energy component, or directly or indirectly from the cosmic neutrino background. In each of these examples, we find that a future version of Dark SRF has the potential to explore a wide range of new parameter space for such models. Finally, we conclude in \Sec{conclusion} and discuss directions for future investigation. A series of appendices is also included, which discusses many of the technical details alluded to throughout the main body.

%%%%%%%%%%%%%%%%%%%%%%%%%%%%%%%%%%%%%%%%%%%%%%%%%%%%%%%%%%%%
\section{Conceptual Overview}
\label{sec:overview}
%%%%%%%%%%%%%%%%%%%%%%%%%%%%%%%%%%%%%%%%%%%%%%%%%%%%%%%%%%%%

Before providing the technical details required to determine the behavior of relativistic plasmas, we begin by giving a brief overview of the models, signals, and experimental setup discussed in this work. We will consider a model of dark QED that includes a dark fermion $\x$ charged under a dark photon $\Ap_\mu$ that is kinetically-mixed with the SM photon $A_\mu$,
\be
\label{eq:Lag1}
\Lag \supset -\frac{1}{4} \, F_{\mu \nu} F^{\mu \nu} -\frac{1}{4} \, F_{\mu \nu}^\p F^{\p \, \mu \nu} + \frac{\eps}{2} \, F_{\mu \nu} F^{\p \, \mu \nu} - A_\mu \, J^\mu - \Ap_\mu \, J^{\p \, \mu}
~.
\ee
Above, $F_{\mu \nu}$ and $F_{\mu \nu}^\p$ are the field-strengths for $A_\mu$ and $A_\mu^\p$, respectively, $\eps$ is the small dimensionless kinetic mixing parameter, $J^\mu$ is the SM electromagnetic current, $J^{\p \, \mu} = e^\p \, \bar{\x} \g^\mu \x$ is the dark current, and $e^\p = \sqrt{4 \pi \alpha^\p}$ is the dark gauge coupling. For simplicity, we will take $\x$ and $\Ap$ to be sufficiently light that they can be well-approximated as massless. To leading order in $\eps \ll 1$, \Eq{Lag1} can be diagonalized by redefining the dark photon field as $\Ap \to \Ap + \eps \, A$. In this basis, it is clear that in addition to its interaction with $\Ap$, $\x$ inherits a small effective ``millicharge" under normal electromagnetism, $\qx \simeq \eps \, e^\p / e$, in units of the standard electric charge $e$, whereas SM currents only couple  to the SM photon. If, alternatively, the $\Ap$ possesses a small mass $\mAp$, then $\x$ is effectively millicharged under normal electromagnetism only on length scales smaller than the Compton wavelength $\sim \mAp^{-1}$~\cite{Berlin:2019uco}. Throughout this work, we will consider $\mAp \ll 10^{-7} \ \eV$, corresponding to millicharge-like interactions on meter and longer length scales.

The strongest bounds on light mCPs come from astrophysical considerations. The millicharge coupling enables copious pair-production of $\x^+$ and $\x^-$ in the solar core, which occurs most efficiently through the decays of SM plasmons. Such processes lead to extra energy loss in the Sun and are bounded to contribute less than $1.5\%$ of the observed solar luminosity from helioseismology and solar neutrino data, translating to an upper limit of $\qx \lesssim 2 \times 10^{-14}$~\cite{Vinyoles:2015khy}. Similar reactions can also modify the stellar evolution of red giant stars, which has been used to derive the updated bound $\qx \lesssim 6 \times 10^{-15}$~\cite{Fung:2023euv}.

\begin{figure*}[t!]
\centering
\includegraphics[width=0.8\linewidth]{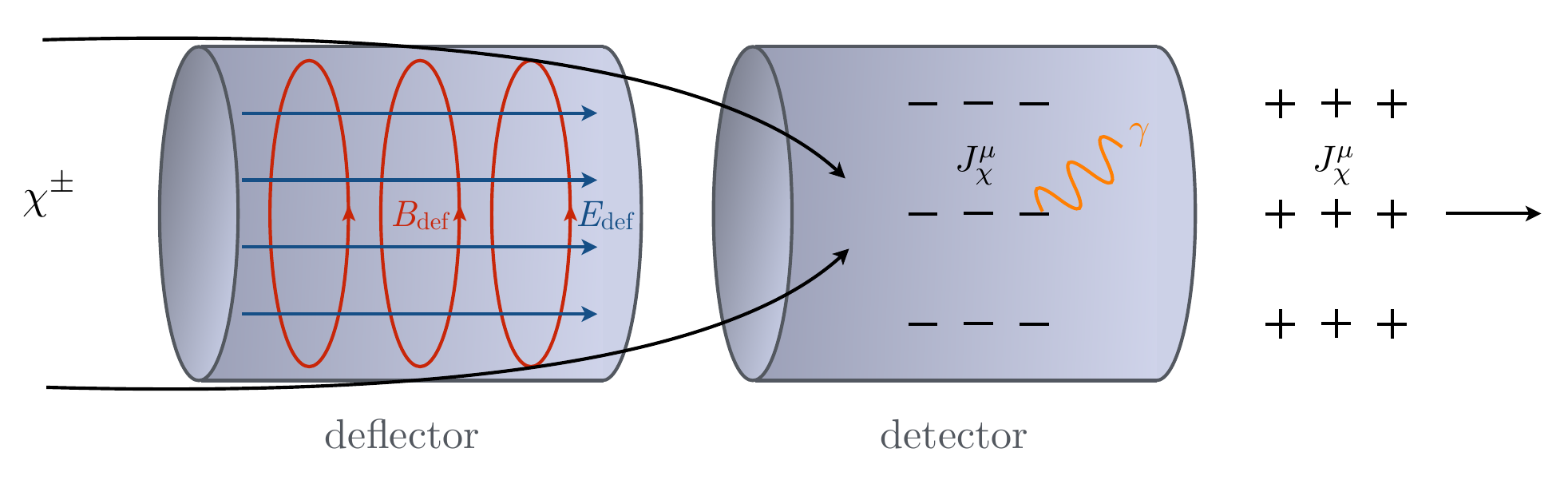}
\caption{A schematic of the ``direct deflection" experimental concept discussed throughout this work. In a typical light-shining-through-wall experiment employing RF cavities, a ``deflector" cavity is driven to high-field. A quiet ``detector" cavity, tuned to the same frequency, is placed nearby. If a feebly-coupled background plasma of millicharged $\x^\pm$ radiation is present, the driven electromagnetic fields of the deflector cavity create disturbances in the plasma in the form of charge and current densities $J_\x^\mu$ that oscillate at the same frequency as the driven cavity. This alternating wave train of millicharges and millicurrents is able to propagate unimpeded throughout the experiment, resonantly exciting electromagnetic fields in the shielded detector cavity.}
\label{fig:cartoon}
\end{figure*}

In this work, we focus on ``direct deflection" experiments, which were proposed in Ref.~\cite{Berlin:2019uco} and further investigated in Ref.~\cite{Berlin:2021kcm} as a laboratory probe of non-relativistic mCPs. Here, we show that analogous setups can also search for a \emph{relativistic} background of low-energy mCPs (of cosmological or astrophysical origin) with sensitivity extending beyond current astrophysical limits. As shown schematically  in \Fig{cartoon}, this detection scheme involves two regions. The first, referred to as the ``deflector," involves strong electric and magnetic fields with amplitudes $E_\text{def}$ and  $B_\text{def}$, extending over a shielded region of length $\Ldef$ and oscillating at frequency $\wdef$. The trajectories of mCPs passing through this region are electromagnetically deflected, setting up small millicharge $\rho_\x$ and millicurrent $\jv_\x$ density perturbations. These perturbations propagate at relativistic speeds into a separate shielded ``detector" region, generating signal electromagnetic fields that oscillate at the same frequency. An electromagnetic detector tuned to $\wdef$ can resonantly amplify such fields, enhancing the sensitivity to small couplings. Such setups thus operate analogously to LSW experiments, where instead of producing and detecting new particles, the system induces and detects disturbances in a background of feebly-interacting particles. 

In the weak-coupling, collisionless, and quasistatic regime (to be discussed further below), the mCP perturbations can be estimated from the force imparted by the deflector $F_\text{def}=eq_\x \left|\E_\text{def}+\vv_\x \times \B_\text{def} \right|\sim e q_\x E_\text{def}$, by noting that the change in the velocity of an mCP with energy $E_\x$ and velocity $v_\x$ is $\delta v_\x \sim e q_\x \, E_\text{def} \, \, \delta t_\text{def} / E_\x$, where $\delta t_\text{def} \sim \Ldef / v_\x$ is the timescale that the mCP experiences a coherent electromagnetic deflection, set by the mCP transit time. This estimate holds in the limit that the deflector is quasistatic, i.e., $\wdef \ll v_\x / \Ldef$. 
The amplitude of the current perturbations induced inside the deflector is thus~\cite{Berlin:2019uco,Berlin:2021kcm}\footnote{Here, we have used the force imparted \emph{transverse} to the incoming direction of the mCP, since the longitudinal component of the acceleration is suppressed by $\g_\x^2$, where $\g_\x$ is boost of an individual mCP. In later sections, we will quantify the relative \emph{bulk} motion of the plasma by $\g$ (without a ``$\x$" subscript). } 
\be
\label{eq:rhoestimate}
|\jv_\x| \sim e q_\x \, n_\x \, \delta v_\x \sim \frac{\wpt^2}{v_\x} \, E_\text{def} \, \Ldef 
~,
\ee
where $\wpt \sim e q_\x \, \sqrt{n_\x / E_\x}$ is the contribution to the plasma frequency of the SM photon from the background of mCPs with number density $n_\x$. Alternatively,  \Eq{rhoestimate} follows from the Drude model for the conductivity $\sigma$ of a collisionless plasma, i.e., $\sigma = |\jv_\x| / E_\text{def} \sim \delta t_\text{def} \, \wpt^2$. 

Note that \Eq{rhoestimate} is akin to the standard result of Debye screening, which states that a weakly-coupled background of charged particles partially screens the deflector's electric potential $A^0_\text{def} \sim E_\text{def} \, \Ldef$, setting up a charge density of $\rho_\x \sim |\jv_\x| / v_\x \sim (\wpt / v_\x)^2 \, A^0_\text{def}$ (see, e.g., Ref.~\cite{Vernet:debye} for a qualitative discussion). In particular, when the plasma is isotropic in the frame of a static deflector, the induced density tracks the local value of the potential, $\rho_\x (\xv) \propto A^0_\text{def} (\xv)$, such that no perturbations exist where the laboratory is electrically grounded. This would seem to imply that no charge densities should propagate out of a shielded deflector.  However, provided that certain criteria are met, $\rho_\x$ and $\jv_\x$ can indeed propagate into a separate shielded ``detector region." For instance, if the deflector (i.e., laboratory) frame is distinct from the rest frame of the plasma, the relative ``wind" of charged particles in the lab frame allows the charge and current perturbations to propagate downstream~\cite{Berlin:2019uco}. Also, on timescales shorter than the transit time $\delta t_\text{def} \sim \Ldef / v_\x$, the behavior of the plasma exhibits transient behavior that deviates from the $\rho_\x (\xv) \propto A^0_\text{def} (\xv)$ steady-state solution. Hence, in either case of an mCP wind or non-quasistatic deflector, the perturbations can penetrate the detector at a level parametrically similar to  \Eq{rhoestimate}. This is shown in detail in the next section, which outlines the formalism to more accurately describe the response of the mCP background, incorporating the complete magnetohydrodynamic response of relativistic mCPs. For instance, as we will see, this treatment accounts for modifications to the simple estimate in \Eq{rhoestimate} that can arise as a result of collective backreactions from long-ranged inter-mCP interactions.

As emphasized above, \Eq{rhoestimate} applies to the quasistatic case where $\wdef \ll v_\x / \Ldef$. For much higher frequencies, we instead expect a strong suppression in $\rho_\x$ and $\jv_\x$, since in this case an mCP does not experience a coherent force as it traverses the deflector. As a result, past studies, which have focused on non-relativistic populations of mCPs with $v_\x \ll 1$, identified LC circuit resonators as the ideal  detector to operate at sub-GHz frequencies~\cite{Berlin:2019uco,Berlin:2021kcm}. In this work, we instead focus on relativistic plasmas, implying that RF cavities with $\wdef \sim 1 / \Ldef$ can be used as both the deflector and detector, analogous to the existing Dark SRF LSW experiment employing superconducting cavities~\cite{Romanenko:2023irv}. 

Also note that from \Eq{rhoestimate} the perturbations scale inversely with the characteristic energy of the plasma, $|\jv_\x| \propto \wpt^2 \propto 1/E_\x$. As a result, the sensitivity of a direct deflection setup is enhanced for lower energy systems. In this work, we will focus on two low-energy examples that arise in a relativistic context: the so-called ``dark solar wind" population of mCPs produced from the Sun~\cite{Chang:2022gcs}, as well as cosmological populations of dark radiation with characteristic temperature $\ll \meV$. Before discussing the sensitivity to each scenario, in the next section we first provide a summary of our formalism, which can be generally applied to ultrarelativistic plasmas.

%%%%%%%%%%%%%%%%%%%%%%%%%%%%%%%%%%%%%%%%%%%%%%%%%%%%%%%%%%%%
\section{Deflecting Dark Radiation}
\label{sec:formalism}
%%%%%%%%%%%%%%%%%%%%%%%%%%%%%%%%%%%%%%%%%%%%%%%%%%%%%%%%%%%%

%%%%%%%%%%%%%%%%%%%%%%%%%%%%%%%%%%%%%%%%%%%%%%%%%%%%%%%%%%%%
\subsection{Plasma Formalism}
%%%%%%%%%%%%%%%%%%%%%%%%%%%%%%%%%%%%%%%%%%%%%%%%%%%%%%%%%%%%

This section discusses the general formalism used to determine the millicharged plasma's response to a driven electromagnetic field and may be skipped for those readers solely interested in its application when determining the sensitivity to concrete examples. In this subsection and in the appendices, we will work with quantities evaluated in the rest frame of the mCP plasma (denoted with tildes). We treat the mCPs and the dark photons collectively as an ultrarelativistic thermal plasma, described by a temperature $\tilde{T}_\x$ much greater than the particle masses (note that the plasma temperature is only well-defined in its rest frame). For instance, $\tilde{T}_\x$ determines the number density as $\tilde{n}_\x = 3\zeta(3)\, g_\x \, \tilde{T}_\x^3 / 4\pi^2 \simeq 0.4 \, \tilde{T}_\x^3$ with $g_\x=4$ for the spin-states of $\x^\pm$~\cite{Thoma:2008my}.

The response of an mCP population to the deflector is well-captured by fluid variables (e.g., charge and current densities) in the limit that there are many mCPs within the deflector. As an example, taking the deflector to be a spherical cavity of radius $\Ldef$ in the laboratory frame and the relative motion of the plasma and the laboratory to be described by the Lorentz factor $\g$, the typical number of mCPs inside the deflector in the plasma frame is then $\tilde{N}_\x \sim \tilde{n}_\x \, (4 \pi / 3) \, \Ldef^3 / \g$. To be well-approximated as a fluid, we will then demand that $\tilde{N}_\x \gtrsim 10^2$, such that the relative size of Poisson fluctuation  in the total mCP number is small, $1 / \tilde{N}_\x^{1/2} \lesssim 0.1$, corresponding to $\tilde{T}_\x \gtrsim 10^{-3} \ \meV \times \g^{1/3} \, (\text{m} / \Ldef)$.\footnote{We stress that here $\tilde{N}_\x$ is defined in the plasma frame. Due to loss of the notion of simultaneity between frames, the number of mCPs that are simultaneously inside the deflector is frame-dependent.}

We will also approximate the plasma response in the weak-field and collisionless limits~\cite{lifschitz1983physical,Blaizot:2001nr}. The plasma is said to be in the weak-field regime when its potential energy is small compared to its kinetic energy $\sim 3 \, \tilde{T}_\x$. The potential energy of the mCP interacting with the deflector is, $e q_\x \, \tilde{E}_\text{def} \, \tilde{L}_\text{def} \sim e q_\x \, \g \, E_\text{def} \, \Ldef$, whereas the typical interaction energy between neighboring mCPs in the plasma is $\alpha^\p \, \tilde{n}_\x^{1/3}$. Thus, for $E_\text{def} \sim 0.1 \ \text{T}$ and $\Ldef \sim 1 \ \text{m}$, the weak-field condition is satisfied for $\tilde{T}_\x \gtrsim 10^{-5} \ \meV \times \g \left(q_\x/ 10^{-14}\right)$ and $\alpha^\p \lesssim 1$. The plasma is in the collisionless regime on timescales smaller than the inverse momentum-exchange rate $\sim \big( \alpha^{\p \, 2} \, \tilde{T}_\x \big)^{-1}$~\cite{Thoma:1994fd,Mrowczynski:1989np,Baym:1997gq}. Therefore, the plasma is collisionless within the time it takes to traverse the deflector provided that $\alpha^{\p \, 2} \, \tilde{T}_\x \lesssim \g / \Ldef$, or equivalently $\alpha^\p \lesssim 10^{-4} \times \g \, (\meV / \tilde{T}_\x) \, (\text{m} / \Ldef)$.

As a toy example, let us consider such a plasma consisting of charged particles $\x^\pm$ that couple only to the SM electromagnetic field. We will therefore ignore in this section the role of the dark photon, but will address this point later. We wish to describe the currents $J_\x^\mu$ in the plasma that are induced as a result of the deflector, the latter of which is described as a stiff source-current $J_\text{def}^\mu$. In the plasma frame, Maxwell's equations are given by 
\be
\label{eq:Maxwell1}
\partial_\mu \tilde{F}^{\mu \nu} (x) = \tilde{J}_\x^\nu (x) + \tilde{J}_\text{def}^\nu (x) 
~.
\ee
In the weak-field limit, the plasma response is approximately linear in the electromagnetic field, such that in momentum-space the current induced in the plasma is
\be
\label{eq:LinResp1}
\tilde{J}_\x^\mu (k) = \tilde{\Pi}^{\mu \nu} (k) \, \tilde{A}_\nu (k) 
~.
\ee
The tensor $\tilde{\Pi}^{\mu \nu}$ can be derived from either the Vlasov (transport) equations~\cite{Silin:1960pya,Klimov:1982bv} or thermal field theory~\cite{Weldon:1982aq} (the two approaches have been shown to give equivalent results~\cite{Kelly:1994dh,Kelly:1994ig,Blaizot:1993zk,melrose2008quantum}). In \App{VlasovDerivation}, we provide a derivation of $\tilde{\Pi}^{\mu \nu}$ using the Vlasov equation for an ultrarelativistic plasma. As shown there, $\tilde{\Pi}^{\mu \nu} (k)$ depends on the plasma frequency $\wpt$ in the plasma rest frame, which is generally defined as \cite{Raffelt:1996wa}
\be
\wpt^2 = (e q_\x)^2 \, \tilde{n}_\x \, \bigg\langle \frac{1-\tilde{v}_\x^2/3}{\tilde{E}_\x} \bigg\rangle
~,
\ee
where the brackets involve an average over phase space of the mCP velocity $\tilde{v}_\x$ and energy $\tilde{E}_\x$ in the plasma rest frame. 

Using \Eq{LinResp1} in the Fourier transform of \Eq{Maxwell1} then yields~\cite{melrose2008quantum}
\be
\label{eq:MaxwellEq}
\big( k^2 \, \eta^{\mu\nu} - k^\mu \, k^\nu +  \tilde{\Pi}^{\mu\nu} \big) \, \tilde{A}_{\nu} (k) = - \tilde{J}_\text{def}^{\mu} (k)
~.
\ee
As discussed in \App{MaxwellCoulombgauge}, the above equation can be solved after decomposing $\tilde{\Pi}^{\mu \nu}$ into its longitudinal $\tilde{\Pi}_L$ and transverse $\tilde{\Pi}_T$ components. In Coulomb gauge, $k^i \, \tilde{A}^i (k)=0$, this procedure gives
\be
\label{eq:A0Ai}
\tilde{A}^{0} (k) = \frac{\tilde{J}_\text{def}^0 (k)}{|\kv|^2 + (|\kv| / \w)^2 \, \tilde{\Pi}_L}
~~,~~
\tilde{A}^i (k) = \frac{\tilde{J}_\text{def}^i (k) - (k^i\w/|\kv|^2) \, \tilde{J}_\text{def}^0 (k)}{|\kv|^2-\w^2-\tilde{\Pi}_T}
~.
\ee
The poles in the above expressions determine the longitudinal and transverse plasma dispersion relations, $\w^2+\tilde{\Pi}_L (k) = 0$ and $\w^2-|\kv|^2+\tilde{\Pi}_T (k) = 0$, respectively. The currents induced in the plasma are then determined by using \Eq{A0Ai} in \Eq{LinResp1} (see also \Eq{Jiind})
\begin{align}
\label{eq:rhoDindtilde}
\tilde{J}_\x^0 (k) &= - \frac{\tilde{\Pi}_L}{\w^2 + \tilde{\Pi}_L} \, \tilde{J}_\text{def}^0 (k)
\\
\label{eq:JDindtilde}
\tilde{J}_\x^i (k) &= \frac{\tilde{\Pi}_T}{|\kv|^2 - \w^2 - \tilde{\Pi}_T} \, \tilde{J}_\text{def}^i (k) - \bigg( \frac{\tilde{\Pi}_L}{\w^2 + \tilde{\Pi}_L} + \frac{\tilde{\Pi}_T}{|\kv|^2 - \w^2 - \tilde{\Pi}_T} \bigg) \, \frac{\w \, k^i}{|\kv|^2} \, \tilde{J}_\text{def}^0 (k)
~.
\end{align}
Finally, we evaluate $\tilde{J}_\x^\mu (x)$ by taking the inverse-Fourier transform of the above momentum-space expressions, which in our case needs to be done numerically (see \App{inverseFourier} for additional details). 

As a simple example, let us take the deflector to consist of a static test charge $q_\text{def}$  at rest in the plasma frame, such that $\tilde{J}_\text{def}^0 (x) = e q_\text{def} \, \delta^3 (\xv)$. From \App{isoplasma}, the static ($\w \to 0$) limit of the longitudinal plasma tensor is $\tilde{\Pi}_L \simeq 3 \, (\wpt \, \w / |\kv|)^2$. Using this in  \Eq{rhoDindtilde} and inverse-Fourier-transforming to position-space then gives the standard result, $\tilde{J}_\x^0 (x) = - 3 \wpt^2 ~ \tilde{A}^0_\text{def} (\xv) ~ \text{exp}(- \sqrt{3} \, \wpt \, |\xv|)$, where $\tilde{A}^0_\text{def} (\xv) = e q_\text{def} / (4 \pi |\xv|)$ is the electric potential of the deflector. 
Thus, within a distance $\sim \wpt^{-1}$ from the deflector, the charge density induced in the plasma agrees with the parametric form given in \Sec{overview}.

The above result, in which case Debye screening gives rise to a plasma charge density that is directly proportional to the local electric potential, occurs in the special case where the deflector is static and the momentum distribution of the plasma is isotropic in the lab frame. However, in our work, we will consider a deflector that is not stationary in the plasma frame, which can arise if it oscillates in time or has a significant relative motion with respect to the plasma. Importantly, the time it takes for the plasma to respond to the deflector is roughly the crossing time of a single particle. As a result, the plasma response continually lags behind that of a rapidly oscillating deflector, screening instead an earlier electromagnetic configuration. This is known as \textit{dynamical Debye screening} \cite{Chu:1988wh,Chakraborty:2006md}. In the next section, we numerically evaluate results for such a deflector in the plasma frame, and then boost into the laboratory frame, in order to determine the induced response for various representative choices of model and experimental parameters.

%%%%%%%%%%%%%%%%%%%%%%%%%%%%%%%%%%%%%%%%%%%%%%%%%%%%%%%%%%%%%%
\subsection{Perturbing a Relativistic Plasma}
\label{sec:perturb}
%%%%%%%%%%%%%%%%%%%%%%%%%%%%%%%%%%%%%%%%%%%%%%%%%%%%%%%%%%%%%%

In this section, we determine the form of the plasma current $J_\x^\mu = (\rho_\x , \jv_\x)$ induced in the laboratory (i.e., deflector) frame. This coordinate system is denoted by symbols without tildes, with the origin defined to be at the center of the deflector. When investigating the effect of the relative motion between the laboratory and plasma frames, we will take the resulting plasma ``wind" in the laboratory frame to be oriented along the $-\hat{z}$ direction. Along these lines, we will dominantly focus on two representative cases, quantified by the boost $\g$ of this wind: $\g = 1$, corresponding to no relative motion between the deflector and plasma, and $\g = 893$, corresponding to a large wind with the particular value of $\g$ motivated by the ``dark solar wind" scenario investigated later in \Sec{darksolarwind}.

As discussed in the previous subsection, the deflector is incorporated by the four-current $J_\text{def}^{\mu}$. As a representative example, we take the deflector in the lab frame to be composed of two separate localized contributions separated by a distance $d$ along the $z$-direction, $J_\text{def}^\mu (x) = \big( J_\text{cap}^\mu (\xv + d \, \zhat/2) + J_\text{cap}^\mu (\xv - d \, \zhat /2) \big) \, e^{-i \wdef t}$, where a contribution of a single piece $J_\text{cap}^\mu (\xv)$ is defined to be
\begin{align}
\label{eq:Jdef}
J_\text{cap}^{0} (\xv) &= \mathcal{J}_\text{def} ~ \frac{z}{\Ldef} \, e^{-z^2/\Ldef^2} \,\cos\left(x / \Ldef\right) \, \cos\left(y / \Ldef\right) 
\nl
J_\text{cap}^{x} (\xv) &= J_\text{cap}^{y} (\xv) = 0
\nl
J_\text{cap}^{z} (\xv) &= - \frac{i \mathcal{J}_\text{def}}{2} \,  \wdef \, \Ldef\, e^{-z^2/\Ldef^2} \, \cos\left(x / \Ldef\right) \, \cos\left(y / \Ldef\right) 
~,
\end{align}
and the amplitude $\mathcal{J}_\text{def} \sim B_\text{def}/\Ldef$ controls the overall strength of the current. Note that the form of $J_\text{cap}^\mu (\xv)$ corresponds to a source that is exponentially localized within a distance $\Ldef$ to the $x-y$ plane at $z = 0$ and is periodic over the same length scale in the transverse $x$ and $y$ directions. Thus, in the static $\wdef \to 0$ limit, $J_\text{cap}^\mu (\xv)$ is  a fixed charge distribution qualitatively similar to a parallel plate capacitor in the $x-y$ plane with a sinusoidal charge profile. The deflector $J_\text{def}^\mu (x)$ is a sum of two such contributions, one localized near $z = - d/2$ and one at $z = d/2$. Our motivation for \Eq{Jdef} is as follows: 1) it obeys charge continuity, 2) it admits a closed-form Fourier transform (see \App{acdef}), and 3) for a particular value of the separation $d$, the electromagnetic fields that $J_\text{def}^\mu$ sources decay exponentially with $|z|$ (analogous to a shielded deflector) and are qualitatively similar to those employed in LSW experiments for $|\xv| \lesssim \Ldef$. This is satisfied by fixing $d = \pi \, \Ldef /  \sqrt{(\wdef \, \Ldef)^2 - 2}$ for $\wdef \, \Ldef > \sqrt{2}$ and $d = 0$ for $\wdef \, \Ldef < \sqrt{2}$.  

\begin{figure*}[t!]
\centering
\includegraphics[width=0.49\linewidth]{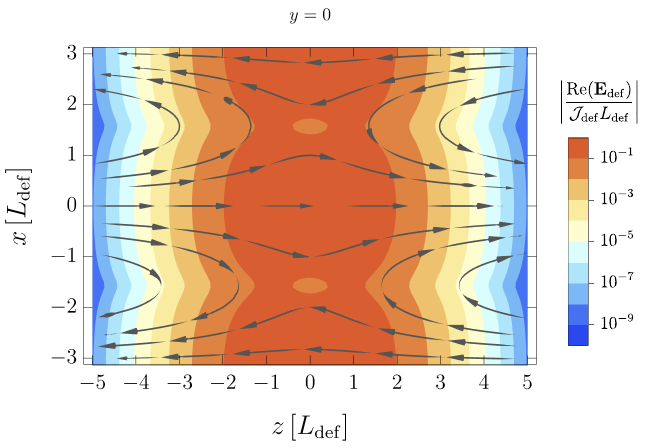}
\includegraphics[width=0.49\linewidth]{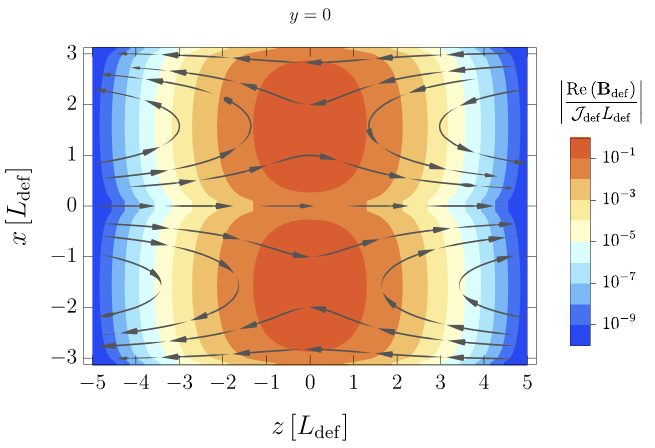}
\caption{The electromagnetic fields $\E_\text{def}$ and $\B_\text{def}$ directly sourced by the deflector of width $\sim \Ldef$ (\Eq{Jdef}) and frequency $\wdef = 2.4 \, \Ldef^{-1}$, independent of the plasma, evaluated  in the $x-z$ plane at $y = 0$ and $t=\Ldef$. In both panels, the dark gray arrows show the direction of $\text{Re} (\E_\text{def})$ projected onto the $x-z$ plane. In the left- and right-panels, the color contours show the magnitude $|\text{Re} (\E_\text{def})|$ and $|\text{Re} (\B_\text{def})|$, respectively, normalized by $\mathcal{J}_\text{def} \, \Ldef$.}
\label{fig:EBdef}
\end{figure*}

To demonstrate this last point, we solve Maxwell's equations with a source given by \Eq{Jdef} in order to determine the electric $\E_\text{def}$ and magnetic $\B_\text{def}$ fields directly generated by the deflector, independent of the plasma. Our results are displayed in \Fig{EBdef}, which shows the direction of $\text{Re} (\E_\text{def})$ in both panels, as well as the magnitude $|\text{Re}(\E_\text{def})|$ and $|\text{Re}(\B_\text{def})|$ in the left- and right-panels, respectively. Here, we fix the deflector frequency to be $\wdef = 2.4/\Ldef$, where the numerical prefactor is motivated by the form of the $\text{TM}_{010}$ mode of a cylindrical cavity of radius $\Ldef$~\cite{Hill}. We find that $E_\text{def} \sim B_\text{def} \sim \mathcal{J}_\text{def} \, \Ldef$ are approximately constant within a distance $\Ldef$ of the origin and are exponentially-suppressed for $|z| \gtrsim \Ldef$ (the precise value of the separation $d \sim \Ldef$ is fixed to guarantee this), with $\E_\text{def}$ dominantly aligned along the $z$-axis for $|\xv|\lesssim \Ldef$. This is qualitatively similar to the cavity-mode structure of the $\text{TM}_{010}$ mode employed in the Dark SRF LSW experiment~\cite{Romanenko:2023irv}. Indeed, we find that when integrated over the volume defined by $\sqrt{|x|^2 + |y|^2} \, , \, |z| \leq \pi \Ldef / 2$, $\E_\text{def}$ and $\B_\text{def}$ have $\order{1}$ overlap with the $\text{TM}_{010}$ modes of a cylindrical cavity aligned along the $z$-direction and centered at $|\xv|=0$.

To determine the induced plasma current $J_\x^\mu (x)$, we first Lorentz boost the deflector current $J_\text{def}^\mu (x)$ to the plasma frame, Fourier transform to momentum-space, use this in \Eqs{rhoDindtilde}{JDindtilde} to obtain $\tilde{J}_\x^\mu (k)$, and then numerically perform an inverse-Fourier transform back to position-space.\footnote{As a consistency check, we have confirmed that our numerically obtained $J_\x^\mu (x)$ obeys charge continuity.} The fact that $J_\text{def}^\mu (x)$  is sinusoidal in $t$, $x$, and $y$ implies that the corresponding Fourier integrals are trivial.  This is an important simplification for our analysis, which makes the highly-oscillatory integrals amenable to numerical evaluation. This same sinusoidal profile, however, implies that the source is infinitely extended along the transverse directions; while this is unrealistic, we expect our results to be accurate up to $\order{1}$ factors as long as we limit ourselves to $|\xv| \lesssim  \Ldef$ since at these locations the fields sourced by $J_\text{def}^\mu$ are qualitatively similar to those of LSW cavity modes, as discussed above.

\begin{figure*}[t!]
\centering
\includegraphics[width=0.543\linewidth]{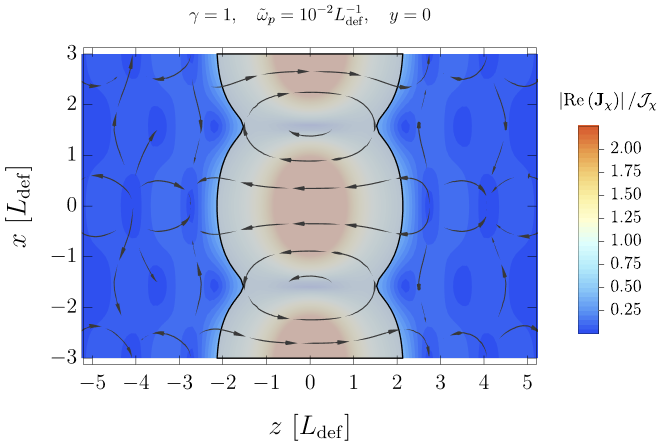}
\includegraphics[width=0.44\linewidth]{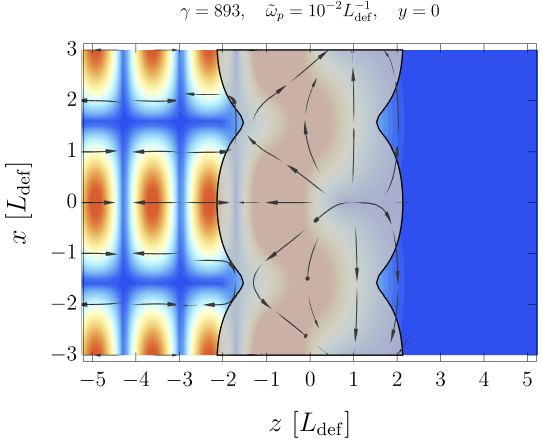}   
\caption{The plasma current $\jv_\x$ in the lab-frame, evaluated in the $x-z$ plane at $y = 0$ and $t=\Ldef$, fixing the plasma frequency to $\wpt = 10^{-2} \, \Ldef^{-1}$ and deflector frequency to $\wdef = 2.4 \, \Ldef^{-1}$. In both panels, the dark gray arrows show the direction of $\text{Re}(\jv_\x)$ projected onto the $x-z$ plane, and the contours show the magnitude $|\text{Re} (\jv_\x)|$, normalized by $\mathcal{J}_\x$ (\Eq{Jxnorm}). In the left-panel, we consider the plasma to be at rest in the lab frame with no boost $\g = 1$, whereas in the right-panel we take it to have a large relative boost $\g  = 893$ along the $-z$ direction (corresponding to the scenario investigated in \Sec{darksolarwind}). 
The boundary of the shaded gray region is where $|\E_\text{def} (\xv)| = 1\%$ of its maximal value, thus defining the region in which the deflector fields  are dominantly localized.}
\label{fig:Jprofile}
\end{figure*}

The induced plasma current oscillates with the same time-dependence as the deflector, $J_\x^\mu (x) = J_\x^\mu (\xv) \, e^{-i \wdef t}$. In \Fig{Jprofile}, we show the spatial profile of the vector-current $\text{Re} ( \jv_\x)$, as well as the magnitude $|\text{Re} ( \jv_\x )|$, in the $x-z$ plane at $t=\Ldef$ and $y=0$, for the case where the plasma is either at rest in the lab frame with no boost $\g = 1$ (left-panel) or has a large relative boost $\g = 893$ along the $-z$ direction (right-panel). In both cases, we also fix  $\wdef = 2.4 / \Ldef$ and $\wpt = 10^{-2} / \Ldef$. From this, we see that for $\sqrt{|x|^2 + |z|^2} \lesssim \Ldef$, $\jv_\x$ is aligned or anti-aligned with the $z$-axis in both panels, which is qualitatively similar to the deflector's electric field $\E_\text{def}$ in \Fig{EBdef}. Furthermore, $\jv_\x$ oscillates spatially with period $\sim \Ldef$  away from the deflector, corresponding to wave trains of alternating charge that propagate outward from the source. For the case of $\g = 1$, these pulses propagate symmetrically along the $\pm z$ directions and are strongly-peaked near the source ($|z| \lesssim \Ldef$), whereas for $\g \gg 1$ they propagate out to much larger distances predominantly downwind along the $-z$ direction. In presenting our results, we have normalized the magnitude of the plasma current by a constant $\mathcal{J}_\x$ that is comparable to the heuristic estimate in \Eq{rhoestimate} with $E_\text{def} \sim \mathcal{J}_\text{def} \, \Ldef$,
\be
\label{eq:Jxnorm}
\mathcal{J}_\x \equiv (\wpt \, \Ldef)^2 \mathcal{J}_\text{def} / 10
~,
\ee
where we have fixed the overall numerical prefactor such that $|\jv_\x| = \order{1} \times \mathcal{J}_\x$ near the deflector. 

\begin{figure*}[t!]
\includegraphics[width=0.49\linewidth]{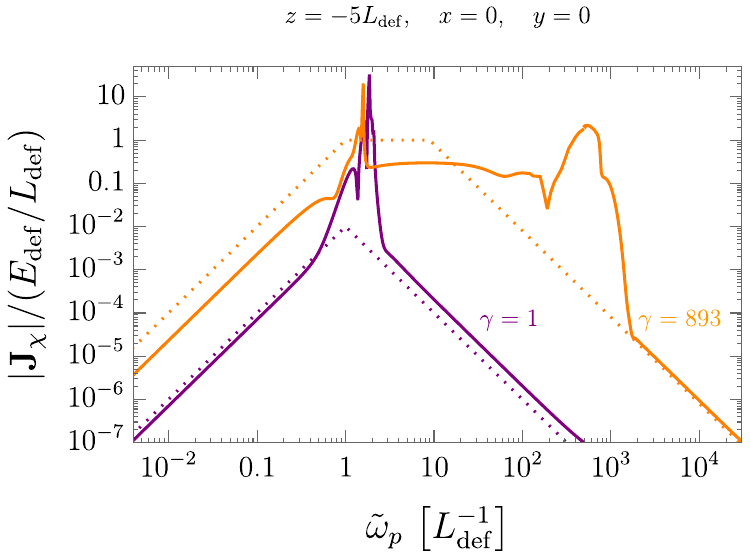}
\includegraphics[width=0.49\linewidth]{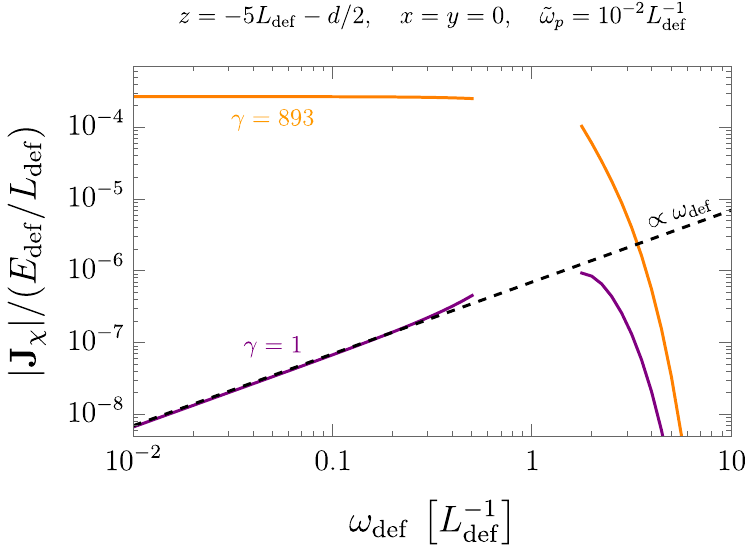}
\caption{The magnitude of the induced plasma current $|\jv_\x|$ in the lab frame, normalized by $E_\text{def} / \Ldef$, where $E_\text{def}$ is the maximum size of the electromagnetic field directly sourced in the interior of the deflector. \textbf{Left}: $|\jv_\x|$ as a function of the plasma frequency $\wpt$ for various boosts $\g$ of the plasma in the lab frame, evaluated at the position $\xv = (0,0, -5 \Ldef)$ and fixing the deflector frequency to $\wdef = 2.4 / \Ldef$. The solid lines correspond to a numerical evaluation, whereas the dotted lines show the simple fitting formula of \Eq{Jxfit}. \textbf{Right}: As in the left-panel, except showing $|\jv_\x|$ as a function of $\wdef$, evaluated at the position $\xv = (0,0, -5 \Ldef - d/2)$, fixing $\wpt = 10^{-2}  / \Ldef$. The separation $d$ is fixed as a function of $\wdef$ following the discussion below \Eq{Jdef}. We do not show our results for $\wdef \sim (0.5 - 2) / \Ldef$, since in this range our prescription for $d$ gives rise to a large unshielded component of $E_\text{def} \sim B_\text{def}$, and so displays features that are not directly relevant when applying our results to a shielded cavity.}
\label{fig:JvsOmega}
\end{figure*}

The $\wpt^2$ scaling of \Eqs{rhoestimate}{Jxnorm} applies to the $\wpt\lesssim \Ldef^{-1}$ and $\wdef \lesssim \text{few} \times \Ldef^{-1}$ regime. To more generally illustrate the behavior of 
$\jv_\x$ with varying plasma and deflector frequencies, we show in \Fig{JvsOmega} the magnitude $|\jv_\x|$ at a fixed position far from the deflector, as a function of either $\wpt$ (left-panel) or $\wdef$ (right-panel), and for two representative choices of the relative boost $\g$ between the plasma and lab frames. Fixing $\wdef \sim 1 / \Ldef$ in the left-panel, we see that for $\wpt \ll \wdef$ the induced current scales with the plasma frequency as $|\jv_\x| \propto \wpt^2$, in agreement with the heuristic discussion of \Sec{overview}. Instead, when $\wpt \gg \g \, \wdef$, we find that the current is suppressed by the large plasma frequency, scaling instead as $|\jv_\x| \propto 1/ \wpt^2$. 

The reason for this turnover near $\wpt \sim \g \, \wdef$ is simple to understand. In the plasma frame, the frequency of the deflector is boosted to $\tilde{\w}_\text{def} \sim \g \, \wdef$. If $\tilde{\w}_\text{def} \lesssim \wpt$, then the deflector is unable to excite on-shell plasmon excitations which propagate out to large distances. 
In this case, in the plasma frame the deflector drags along with it a cloud of charged particles of size limited by the Debye length $\sim \wpt^{-1}$. Boosting back into the proper frame of this cloud (i.e., the rest frame of the deflector), this length scale is \emph{inverse}-Lorentz-contracted along the longitudinal direction to $\g \, \wpt^{-1}$. Thus, if $\g \, \wpt^{-1} \lesssim \Ldef$, then the plasma backreacts and efficiently screens the effect of the deflector. In other words, $\jv_\x$ becomes suppressed in the large $\wpt$ limit once $\wpt \gtrsim \g \, \w_\Def$ and $\wpt \gtrsim \g / \Ldef$ (which for our choice of parameters are roughly equivalent criteria, since $\wdef \sim 1/\Ldef$).

In the right-panel of \Fig{JvsOmega}, we instead fix the plasma frequency to be small ($\wpt \ll 1 / |z| \ll 1 / \Ldef$) and show $|\jv_\x|$ as a function of the deflector frequency $\wdef$. We see that regardless of the boost $\g$, $\jv_\x$ is suppressed for rapidly oscillating deflectors, $\wdef \gg \Ldef^{-1}$. As discussed in \Sec{overview}, this is due to the fact that the oscillation timescale $\sim \wdef^{-1}$ is shorter than the transit time of a relativistic plasma particle $\sim \Ldef$, such that it is not coherently deflected by the electromagnetic field. Instead, we see that for a quasistatic deflector $\wdef \ll \Ldef^{-1}$, the size of the induced current is strongly suppressed if $\g = 1$, corresponding to no relative motion between the plasma and lab frames; as discussed in \Sec{overview}, this is due to the fact that for $\g = 1$ and $\wdef \ll \Ldef^{-1}$, the leading order response of the plasma tracks the local value of the electric potential. If away from the deflector the local value of the electric potential is small, the leading contribution enters at $\order{\wdef \, \Ldef}$, yielding $|\jv_\x| \sim \wdef \, \Ldef \, \mathcal{J}_\x$. 

To summarize, our results show that for a relativistic plasma the optimal choice for $\wdef$ (i.e., one that enhances $|\jv_\x|$ independent of $\g$) is $\wdef \sim \Ldef^{-1}$, analogous to a resonant cavity. In this case, $\wdef$ is large enough to generate a signal that propagates out of the deflector independent of $\g$, but not so large that the deflector oscillates many times within the transit time of a single plasma particle. In the next section, we will consider the sensitivity of such a setup. In this case, we find that at a distance $\sim \text{several} \times \Ldef$ downwind from the deflector, our results can be roughly approximated by 
\be
\label{eq:Jxfit}
|\jv_\x| \sim a_\g \, \frac{E_\text{def}}{\Ldef} ~ \min \bigg(\wpt \, \Ldef \, , \, 1 \, , \, \frac{b_\g \, \g}{\wpt \, \Ldef} \bigg)^2 \, e^{-i \wdef t}
~,
\ee
where $E_\text{def} \sim B_\text{def}$ refers to the maximum amplitude of the electromagnetic field evaluated inside the deflector cavity, and $a_\g$ and $b_\g$ are defined to be $a_\g \equiv 10^{-2}$, $b_\g \equiv 1$ for $\g \sim 1$ and $a_\g \equiv 1$, $b_\g \equiv 10^{-2}$ for $\g \gg 1$. \Eq{Jxfit} is shown as dotted lines in the left-panel of \Fig{JvsOmega}, which demonstrates that for most values of $\wpt$ it adequately captures the main qualitative behavior of the response, up to $\order{1}$ factors. We note that this fit does not capture the significant signal enhancement evident in our numerical estimate for $\wpt \sim \g \, \wdef \sim \g \, \Ldef^{-1}$, which is due to a coherent deflection of the traversing mCPs for our particular choice of the deflector current $J_\text{def}^\mu$.  

Before proceeding, we remind the reader that this calculation was performed assuming that the plasma particles $\x$ only couple to the SM photon. However, as discussed above, we are interested in the possibility that these interactions are mediated indirectly via an ultralight kinetically-mixed dark photon $\Ap$. In this case, $\x$ couples directly to $\Ap$, and indirectly to electromagnetism with an effective millicharge on distance scales smaller than the dark photon Compton wavelength $\sim \mAp^{-1}$. As a result, self-interactions mediated by the $\Ap$ can backreact on current densities induced in the millicharged plasma when $\wpt^\p \gtrsim \g \, \max (\wdef \, , \, 1/\Ldef)$, where $\wpt^\p = e^\p \, \wpt /  (e q_\x)$ is the contribution of $\x$ to the $\Ap$ plasma frequency. In this case, provided that the dark photon is longer-ranged than the experimental setup ($\mAp \ll \Ldef^{-1}$), the millicharged plasma generates an effective visible millicurrent of nearly the same form as in \Eq{Jxfit}, after making the replacements $\wpt \to \wpt^\p$ and $J_\x \to \eps^2 \, J_\x$,
\be
\label{eq:Jxfit2}
|\jv_\x (x)| \sim a_\g \, \eps^2 \, \frac{E_\text{def}}{\Ldef} ~ \min \bigg(\wpt^\p \, \Ldef \, , \, 1 \, , \, \frac{b_\g \, \g}{\wpt^\p \, \Ldef} \bigg)^2 \, e^{-i \wdef t}
~,
\ee
which maintains the expected scaling $|\jv_\x| \propto \eps^2 \, \wpt^{\p \, 2} = \wpt^2$ in the limit of small coupling.

%%%%%%%%%%%%%%%%%%%%%%%%%%%%%%%%%%%%%%%%%%%%%%%%%%%%%%%%%%%%
\subsection{Experimental Reach}
\label{sec:genreach}
%%%%%%%%%%%%%%%%%%%%%%%%%%%%%%%%%%%%%%%%%%%%%%%%%%%%%%%%%%%%

The previous section determined the form of the current densities $\jv_\x$ induced in the millicharge plasma from an oscillating electromagnetic field, such as those driven in resonant cavities (see \Eq{Jxfit2}). Next, we review how $\jv_\x$ can in turn excite small electromagnetic fields in a nearby shielded cavity, which we refer to as the detector. More concretely, we will focus on a scenario analogous to the existing Dark SRF LSW experiment, which has recently conducted a pathfinder run~\cite{Romanenko:2023irv} employing two resonant cavities tuned to the same frequency. One such cavity is driven at high power, with the goal of directly producing new particles, such as dark photons, that can resonantly excite small electromagnetic fields in a shielded detector cavity tuned to the same frequency. Although inadvertent, this same experiment also operates as a direct deflection setup, with the driven cavity functioning as the deflector for a background of relativistic mCPs. 

Dark SRF employs $\text{TM}_{010}$ modes in both cavities, which are longitudinally aligned with respect to the directions of their electric field profiles, $\E_\text{def}$ and $\E_\text{det}$. In the previous section, we saw that a similar electromagnetic field configuration can also source a millicharge current density $\jv_\x$ with a  spatial profile similar to that of a $\text{TM}_{010}$ mode. The ability for this millicurrent to excite the same mode $\E_\text{det}$ in the \emph{detector} cavity is dictated by the resonant form of the signal power~\cite{Hill},
\be
\label{eq:Psig}
P_\text{sig} = \frac{Q}{\wdef} \, \frac{\big| \int d^3 \xv ~ \jv_\x \cdot \E_\text{det}^* \big|^2}{\int d^3 \xv ~ |\E_\text{det}|^2}
\equiv \frac{Q}{\wdef} \, \eta^2 \, |\jv_\x|^2 \, V_\text{det}
~,
\ee
where $Q$ is the quality factor of the detector cavity, the integrals are over the volume $V_\text{det}$ of the detector cavity, and in the second equality $|\jv_\x|$ is the characteristic amplitude of the induced millicurrent given in \Eq{Jxfit2}. In the second equality, we have also defined the dimensionless overlap form factor $\eta$, which is $\order{1}$ in the case where the spatial profiles of $\jv_\x$ and $\E_\text{det}$ are optimally matched. In our estimates, we will adopt $\eta = 1$.

Let us first consider the existing sensitivity of the recent Dark SRF pathfinder run, adopting  representative experimental parameters of $\wdef = 2 \pi \times \GHz$, $Q = 3 \times 10^{10}$, $E_\text{def} = 5 \ \MV / \text{m}$, and $V_\text{det} = \pi \, (10 \ \cm)^3$. After a total data-taking time of a few hours, this run observed no excess power in the detector cavity above thermal noise, setting an upper bound of $P_\text{sig} \lesssim 3 \times 10^{-16} \ \text{W}$, which was limited by an unwanted frequency offset between the two cavities. Equating this to \Eq{Psig} gives an existing sensitivity of $|\jv_\x| \simeq 5 \times 10^{-13} \ \text{A} / \text{m}^2$. Future runs of Dark SRF are expected to significantly enhance this sensitivity with better frequency matching, larger fields and quality factors,  reduced noise temperatures, and an optimized signal analysis. 

Therefore, to estimate the ultimate reach of a future experiment, we adopt the following upgraded experimental parameters: $Q = 10^{12}$, $E_\text{def} = 60 \ \MV / \text{m}$, and deflector/detector volumes as large as $V_\text{def} = V_\text{det} \sim 1 \ \text{m}^3$. We also assume that the deflector cavity's phase is actively monitored, which enables an optimized signal analysis with a corresponding signal-to-noise ratio given by $\text{SNR} = P_\text{sig} \, t_\text{int} / T_\text{det}$, where $t_\text{int} = 1 \ \text{yr}$ is the total observation time and $T_\text{det} = 10 \ \text{mK}$ is the temperature of the detector cavity~\cite{Graham:2014sha}. Demanding that $\text{SNR} \gtrsim 1$ then gives a future sensitivity to millicurrents of size $|\jv_\x| \simeq 3 \times 10^{-24} \ \text{A} / \text{m}^2 \times (1 \ \text{m} / \Ldef)^2$, where we took $V_\text{def} = V_\text{det} = \pi \, \Ldef^3$ and $\wdef = 2.4 / \Ldef$ (corresponding to a cylindrical cavity of radius and length $\Ldef$). 

To summarize, we will adopt the following sensitivity to $|\jv_\x|$ for the existing pathfinder or future reach of Dark SRF,
\be
\label{eq:Jxreach}
|\jv_\x| \sim 
\begin{cases}
5 \times 10^{-13} \ \text{A} / \text{m}^2
& \text{(existing pathfinder)}
\\
3 \times 10^{-24} \ \text{A} / \text{m}^2 \times (1 \ \text{m} / \Ldef)^2
& \text{(future)}
~.
\end{cases}
\ee
In the sections below, we apply the second line of \Eq{Jxreach} to determine Dark SRF's ultimate sensitivity to two scenarios involving relativistic backgrounds of mCPs. First in \Sec{darksolarwind}, we will consider astrophysical mCPs that are produced from the Sun. For sufficiently large self-couplings $\alpha^\p$, such particles can thermalize through self-interactions in the solar interior, forming a so-called ``dark solar wind," as recently investigated in Ref.~\cite{Chang:2022gcs}. Then, in \Sec{cosmo}, we will consider cosmological sources of mCPs, such as those generated from a thermal bath in the early universe or at much later times through a dynamical dark energy component or dark matter decay/annihilation. We will also investigate models in which the lightest SM neutrino has a small effective millicharge or neutrinos equilibrate with a light millicharged sector after neutrino-photon decoupling, such that the cosmic neutrino background directly or indirectly gives rise to the class of signals discussed throughout this work, respectively.

%%%%%%%%%%%%%%%%%%%%%%%%%%%%%%%%%%%%%%%%%%%%%%%%%%%%%%%%%%%%
\section{Dark Solar Wind}
\label{sec:darksolarwind}
%%%%%%%%%%%%%%%%%%%%%%%%%%%%%%%%%%%%%%%%%%%%%%%%%%%%%%%%%%%%

Light mCPs can be created in large numbers from rare thermal processes in the extreme environment of the solar interior. It is commonly assumed that such particles simply free-stream out of the Sun unscathed, each carrying away energy comparable to the temperature of the solar core $\sim \keV$. While this is often the case, self-interactions mediated by the dark photon can drastically alter this scenario, as recently highlighted in Ref.~\cite{Chang:2022gcs}. The degree to which this is true is determined both by the millicharge $q_\x$, which sets the initial density of mCPs produced in the solar core, and the dark fine-structure constant $\alpha^\p$, which controls the strength of self-interactions. In particular, for 
\be
\label{eq:alphaDSW}
\alpha^\p \gtrsim 5 \times 10^{-6} \times \big( 10^{-14} / q_\x \big)^{1/2}
~,
\ee
number-changing processes (such as $\x\x\rightarrow \x\x \Ap$) become highly efficient, leading to local thermalization of the mCP-$\Ap$ plasma in the Sun~\cite{Chang:2022gcs}. As a result, the mean free path of the mCPs is drastically shortened, causing them to behave collectively as a fluid on solar length scales.

Before number-changing processes become efficient, the phase space of the initial free-streaming mCPs is far from thermal, with a number density very small compared to that of a thermal population with the same characteristic energy per particle. Thermalization correspondingly enhances the number density at the expense of lowering the typical mCP energy. This thermalized plasma behaves as an adiabatically expanding fluid driven by its own thermal pressure, accelerating radially outward to increasingly ultrarelativistic bulk velocities as it moves away from the Sun. The resulting steady-state outflow, referred to as the ``dark solar wind"~\cite{Chang:2022gcs}, is orders of magnitude more dense and less energetic ($\ll \keV$ per particle) compared to a free-streaming population of the same luminosity. 

The underlying fluid equations governing the evolution of the dark solar wind were solved in Ref.~\cite{Chang:2022gcs}. Here, we simply quote the results. Evaluated at Earth, the boost describing the relative bulk motion of the fluid is $\g \simeq 893$, while the temperature and number density in the plasma frame are approximately
\be
\tilde{T}_\x \simeq 10^{-4} \ \eV \times \big( q_\x / 10^{-14} \big)^{1/2}
~~,~~
\tilde{n}_\x \simeq 150 \ \cm^{-3} \times \big( q_\x / 10^{-14} \big)^{3/2}
~.
\ee
In turn, these quantities imply a lab-frame per-particle energy and plasma-frame dark plasma frequency of
\be
\label{eq:wpDSW}
E_\x \simeq 0.5 \ \eV \times \big( q_\x / 10^{-14} \big)^{1/2}
~~,~~
\wpt^\p \simeq 1 \ \GHz \times \big( q_\x / 10^{-14} \big)^{1/2} \, \big( \alpha^\p / 10^{-5} \big)^{1/2}
~.
\ee
For the largest viable couplings, $q_\x \sim 10^{-14}$, the resulting energy flux of mCPs is six orders of magnitude greater than the local kinetic energy flux of galactic DM. Despite this, the dark solar wind is challenging to detect in conventional underground dark matter direct detection experiments searching for elastic scattering. This is because the typical energy exchanged in an mCP-electron collision is $\sim \alpha \, p_\x \sim \meV \times (q_\x / 10^{-14})^{1/2}$, well below the thresholds of existing and future sensors. However, note that the direct deflection signals discussed in \Sec{genreach} benefit from the small energy of the plasma. Indeed, $\wpt^\p$ is enhanced both by the larger number density and smaller mCP energy seeded by thermalization. For the largest viable values of $q_\x$, both effects together enhance $\wpt^\p$ by roughly a factor of $\sim 10^3$ compared to a non-thermalized free-streaming population produced by the Sun.\footnote{This comparison involves evaluating the plasma frequency in the rest frame for the dark solar wind and the laboratory frame for the free-streaming case. Hence, this is valid up to $\order{1}$ factors, since the plasma frequency is controlled by the Lorentz invariant ratio $n_\x / E_\x$.} Since the millicurrent $J_\x$ induced by a deflector scales as $\wpt^{\p \, 2}$ in the weak-coupling limit (as in \Eq{Jxfit}), thermalization of solar mCPs enhances the class of signals detectable with a direct deflection experiment. 

Before applying the formalism of \Sec{formalism}, we note that an additional complication arises from the fact that the same self-interaction processes that drive the dark solar wind population towards a thermal distribution also damp the perturbations induced by the deflector. In particular, \Sec{formalism} applies strictly to plasmas that are collisionless on length scales comparable to the size of the deflector-detector setup. The typical distance traversed between collisions by an mCP of the dark solar wind is $\sim \g \, / \,  (\alpha^{\p \, 2} \, \tilde{T}_\x ) \simeq 2 \ \text{m} \times (1 / \alpha^\p)^2 \, (10^{-14} / q_\x)^{1/2}$, which is longer than the size of an experiment in the parameter space of interest. Thus, on laboratory length scales, we may approximate the dark solar wind as collisionless. 

\begin{figure*}
\centering
\includegraphics[width=0.5\linewidth]{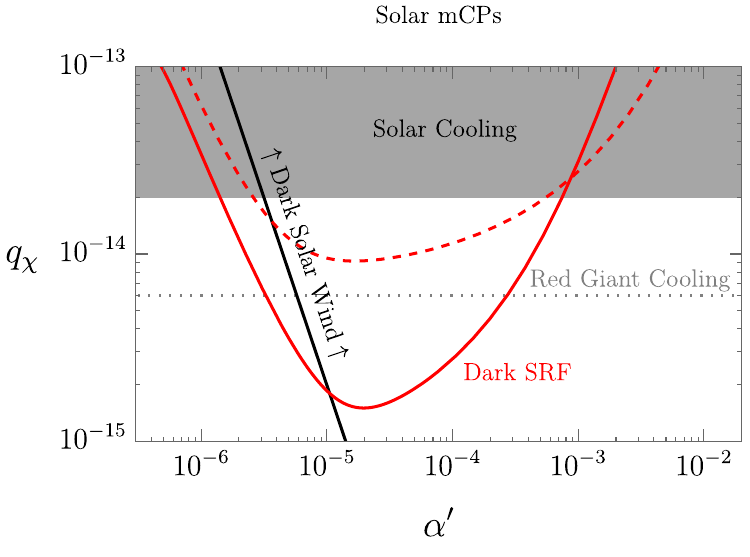}
\caption{The projected direct deflection sensitivity of the Dark SRF light-shining-through-wall experiment to millicharged particles emitted by the Sun, in the plane spanned by the millicharge $q_\x = \eps \, e^\p / e$ and self-coupling $\alpha^\p = e^{\p \, 2} / 4 \pi$, where $\eps$ is the kinetic mixing parameter and $e^\p$ is the dark photon coupling. The solid or dashed red lines assume an experimental setup of size $\Ldef = 1 \ \text{m}$ or $\Ldef = 10 \ \cm$, respectively. Also shown in gray are existing constraints derived from considerations of stellar energy loss~\cite{Vinyoles:2015khy, Fung:2023euv}. We show the Red Giant bound as a dotted line, since the robustness of these constraints has been recently called into question in Ref.~\cite{Dennis:2023kfe} (although see the discussion in Refs.~\cite{Fung:2023euv,Caputo:2024oqc}). Above the black line, millicharged particles thermalize in the solar interior through self-interactions mediated by the dark photon, leading to an enhanced signal in direct deflection experiments.}
\label{fig:darksolarwind}
\end{figure*}

To determine the future sensitivity of the Dark SRF LSW experiment to the dark solar wind, we apply the general results of \Eqs{Jxfit2}{Jxreach}. Note that since \Eq{Jxfit2} applies to regions downwind of the deflector, our projections assume that the axis connecting the two cavities in a LSW experiment is aligned with the Earth-Sun axis. In \Fig{darksolarwind}, we show the ultimate sensitivity of Dark SRF in the $q_\x - \alpha^\p$ plane, fixing the kinetic mixing parameter in terms of the two other couplings, $\eps = e q_\x / \sqrt{4 \pi \alpha^\p}$. 

The sensitivity to $q_\x$ is a non-trivial function of $\alpha^\p$. For $\alpha^\p$ below the critical value in \Eq{alphaDSW}, the mCP population does not efficiently thermalize and simply free-streams out of the Sun. In this case, we determine the dark plasma frequency $\wpt^2 \sim 4 \pi \alpha^\p n_\x / E_\x$ at Earth using $n_\x \simeq ( L_\x / 4 \pi r_\oplus^2 ) / E_\x$, where $E_\x \sim 5 \ \keV$ is the typical mCP energy and the mCP luminosity $L_\x$ is related to the total solar luminosity $L_\odot$ by $L_\x \simeq 4 \times 10^{-2} ~ L_\odot \, (q_\x/10^{-14})^2$~\cite{Vinyoles:2015khy,Chang:2019xva,Chang:2022gcs}. Instead, for $\alpha^\p$ slightly above the critical value in \Eq{alphaDSW}, the dark solar wind develops, and the mCP plasma does not efficiently backreact within the length scales and timescales set by the experiment. For much larger $\alpha^\p$, the backreaction from collective mCP self-interactions screens the perturbations induced by the deflector before reaching the detector cavity, thereby suppressing the signal. As discussed in \Sec{perturb}, such backreactions occur when $\wpt \gg \g / \Ldef$. We find that the resulting range of self-couplings for which a LSW experiment has optimal sensitivity to the dark solar wind is approximately
\be
\label{eq:alphaDSW2}
10^{-5} \times \big( 10^{-14} / q_\x \big)^{1/2} \lesssim \alpha^\p \lesssim 10^{-4} \times \big( 10^{-14} / q_\x \big) \, \big( 1 \ \text{m} / \Ldef \big)^2
~.
\ee

In \Fig{darksolarwind}, we consider two possible sizes of the cavities, $\Ldef = 10 \ \cm$ and $\Ldef = 1 \ \text{m}$, and correspondingly fix the separation of the cavities to be $\sim \text{few} \times \Ldef$ and the operating frequency to be $\wdef = 2.4 / \Ldef$. From \Eq{Jxreach}, larger setups can probe smaller values of the millicharge $q_\x$ in the weak-self-coupling regime. However, since the experimental size and cavity oscillation period decrease as $\Ldef$ is reduced, it is more difficult for the mCP plasma to backreact for smaller values of $\Ldef$. As a result, smaller experimental setups can probe larger values of the self-coupling $\alpha^\p$, as given by \Eq{alphaDSW2}. Regardless, in either case, future iterations of Dark SRF can be sensitive to a wide range of mCP parameter space.

%%%%%%%%%%%%%%%%%%%%%%%%%%%%%%%%%%%%%%%%%%%%%%%%%%%%%%%%%%%%
\section{Cosmological Dark Radiation}
\label{sec:cosmo}
%%%%%%%%%%%%%%%%%%%%%%%%%%%%%%%%%%%%%%%%%%%%%%%%%%%%%%%%%%%%

A relativistic background of mCPs can also be sourced cosmologically, resulting in an approximately isotropic population of millicharged dark radiation. The present abundance of dark radiation is generally quantified by normalizing its present day energy density by the critical energy density, $\ODR = \rho_\text{DR} / \rho_\text{crit}$. In the case of dark radiation in the early universe, $\rho_\text{DR}$ is alternatively parametrized in terms of the additional effective number of neutrino species, $\Delta N_\text{eff} = (8/7) \, (11/4)^{4/3} \, (\rho_\text{DR} / \rho_\g)$, such that $\ODR$ can be reexpressed as $\ODR \simeq 1.2 \times 10^{-5} \, \Delta N_\text{eff}$. An additional cosmological population of dark radiation, beyond the contribution of $N_\text{eff} \simeq 3$ by the cosmic neutrino background, is constrained by Planck and ACT observations of the cosmic microwave background to contribute $\Delta N_\text{eff} \lesssim \order{0.1}$, with the precise value of this upper bound depending on whether such radiation is free-streaming or fluid-like~\cite{Planck:2018vyg,ACT:2020gnv,Blinov:2020hmc}. 

The simplest scenario arises when dark radiation is described by a thermal distribution with temperature $\tilde{T}_\x$. In this case, the resulting energy density and dark plasma frequency are $\tilde{\rho}_\text{DR} = (\pi^2 / 30) \, g_*^\p \, \tilde{T}_\x^4$ and $\wpt^\p = e^\p \, \tilde{T}_\x / 3$, respectively. Here, $g_*^\p$ is the total effective relativistic degrees of freedom in the dark sector, which is $g_*^\p = 11/2$ for a single pair of fermionic mCPs and a  dark photon. The dark plasma frequency can then be related straightforwardly to $\ODR$ by
\be
\label{eq:wpDR}
\wpt^\p \simeq 
93 \ \GHz \times \big( e^\p / g_*^{\p \, 1/4} \big) \,  \big( \ODR / 10^{-5} \big)^{1/4} 
~.
\ee
In this section, we will consider various cosmological sources for $\ODR$. An irreducible contribution arises from the out-of-equilibrium decay of SM plasmons in the early universe~\cite{Dvorkin:2019zdi,Berlin:2022hmt}. In this case, the corresponding density depends directly on the size of the millicharge; $\ODR \simeq 2.5 \times 10^{-16} \times \big( q_\x / 10^{-14} \big)^2$~\cite{Berlin:2022hmt}. Millicharged radiation can also arise from dark matter decay or annihilation. For instance, current limits allow as much as $\sim 4\%$ of the present dark matter density having decayed into dark radiation~\cite{DES:2020mpv}, corresponding to $\ODR \lesssim 10^{-2}$. For dark matter annihilations to mCPs, let us consider the galactic center, which contributes a local millicharge density of $\ODR \sim 10^{-4} \times (\MeV / m_{_\text{DM}}) \, (\sigma v / 10^{-26} \ \cm^3 \ \text{s}^{-1} )$, where $m_{_\text{DM}}$ is the dark matter mass, $\sigma v$ is its annihilation rate to mCPs, and we have taken a $J$-factor of $\sim 10^{24} \ \GeV^2 / \cm^5$ for the inner region of the Milky Way~\cite{Safdi:2022xkm}. Recent work has also investigated radiation sourced by the kinetic energy of a dynamical dark energy component, leading to $\ODR \lesssim 3 \times 10^{-2}$~\cite{Berghaus:2020ekh,Ji:2021mvg,Berghaus:2023ypi}. The population of millicharged radiation from any of these sources initially possesses a non-thermal distribution. However, thermalization can easily occur through self-interactions, analogous to the discussion in \Sec{darksolarwind}. For these examples, we assume this to be the case, and utilize \Eq{wpDR} to determine the corresponding plasma frequency. 

As another example, we consider the possibility that the lightest SM neutrino $\nu$ of the cosmic neutrino background possesses a small effective millicharge, which can arise in scenarios similar to the ``portalino" models of Refs.~\cite{Fox:2011qd,Schmaltz:2017oov}. In this case, electroweak symmetry breaking mixes the active neutrino $\nu$ with a sterile neutrino $N$. If there also exists a dark fermion $N^\pm$ directly charged under a dark $U(1)^\p$, then spontaneous symmetry breaking of the $U(1)^\p$ can mix $N$ and $N^\pm$, thereby giving $\nu$ a small dark charge. If the $\Ap$ of this $U(1)^\p$ kinetically-mixes with SM electromagnetism, then on length scales smaller than $\mAp^{-1}$ the SM-like neutrino inherits a small effective millicharge, suppressed by the size of the kinetic mixing as well as the $\nu - N$ and $N - N^\pm$ mass-mixings. Since a single cosmic neutrino contributes $N_\text{eff} = 1$, this scenario corresponds to $\ODR \simeq 1.2 \times 10^{-5}$. There exist stringent limits on a neutrino millicharge~\cite{Giunti:2014ixa,Giunti:2015gga,Giunti:2024gec}. The strongest of these are model-dependent. For instance, limits derived from tests of matter-neutrality assume that the millicharge is unbroken in the four-Fermi $n \leftrightarrow p e \bar{\nu}_e$ interaction~\cite{Raffelt:1999gv}. However, this need not be the case when the effective millicharge is generated by a broken $U(1)^\p$. Furthermore, astrophysical limits derived from the coupling of neutrinos to magnetic fields~\cite{Barbiellini:1987zz} do not apply if the effective range of the interaction, set by $\mAp^{-1}$, is small compared to astrophysical length scales. In this work, we adopt the most model-\emph{independent} of these limits, which is set by considerations of stellar energy loss~\cite{Vinyoles:2015khy, Fung:2023euv}. We postpone a more complete investigation of these models to future studies. 

The cosmic neutrino background may also \emph{indirectly} contribute to millicharged radiation. In particular, a related scenario arises if the cosmic neutrino background equilibrates with a distinct relativistic mCP population well after neutrino-photon decoupling~\cite{Green:2017ybv,Berlin:2017ftj,Berlin:2018ztp}. If the initial pre-equilibrated energy density of the mCPs is negligible compared to that of the SM radiation bath, then $N_\text{eff} \simeq 3$ is initially unmodified compared to its standard value. When the mCPs thermalize with the SM neutrinos, the temperature of the mCPs increases, thereby lowering the temperature of the neutrinos compared to that of the photon bath. Since equilibration conserves energy, the $\nu + \text{mCP}$ population still contributes only $N_\text{eff} \simeq 3$ after equilibration, with the relative energy density in the mCPs controlled by the ratio of relativistic degrees of freedom $g_*^\p / g_*^\nu$, where $g_*^\nu = 21/4$. Hence, for $g_*^\p \gg g_*^\nu$, the majority of the apparent energy density in the cosmic neutrino background is instead made up of mCPs at late times, analogous to the ``neutrinoless universe" investigated in Ref.~\cite{Beacom:2004yd}. 

\begin{figure*}
\flushleft
\includegraphics[width=0.49
\linewidth]{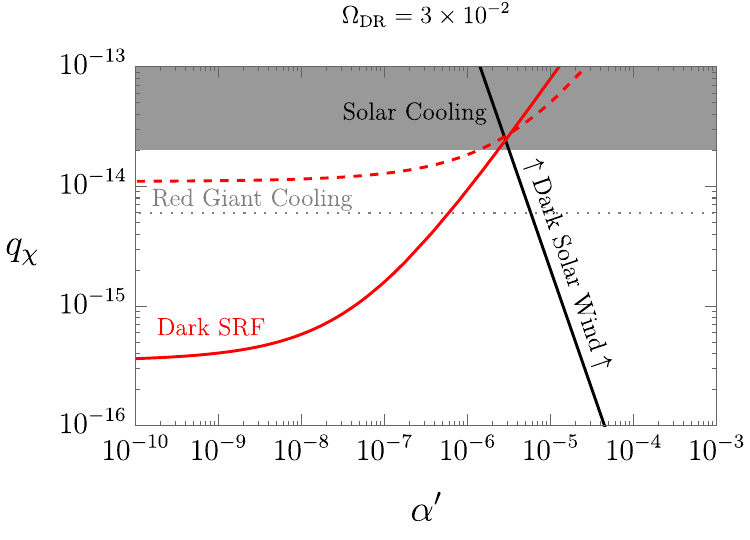}
\includegraphics[width=0.475\linewidth]{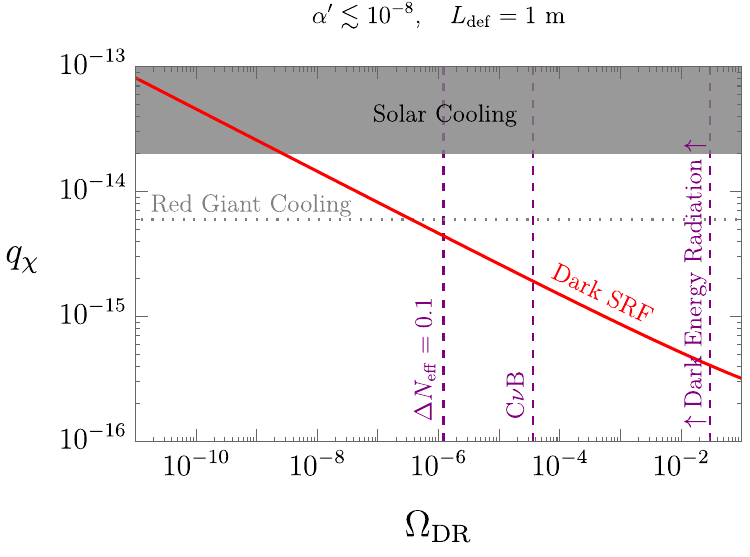}
\caption{As in \Fig{darksolarwind}, but instead for cosmological millicharged dark radiation. \textbf{Left}: Sensitivity of Dark SRF in the $q_\x - \alpha^\p$ plane, for experimental setups of size $\Ldef = 1 \ \text{m}$ (solid red) or $\Ldef = 10 \ \cm$ (dashed red), fixing the density parameter of millicharged radiation to $\ODR = 3 \times 10^{-2}$, corresponding to millicharges sourced by a dynamical dark energy component. Above the black line, the dark solar wind of \Sec{darksolarwind} contributes an irreducible local abundance of millicharged radiation. \textbf{Right}: Sensitivity of Dark SRF in the $q_\x - \ODR$ plane, for an experimental setup of size $\Ldef = 1 \ \text{m}$, fixing the self-coupling to be sufficiently small such that backreactions from self-interactions are negligible on laboratory length scales. As benchmark values of $\ODR$, we show scenarios where millicharged radiation is sourced by dark energy, the cosmic neutrino background (C$\nu$B), or when it is a small subcomponent of the total radiation energy density with $\Delta N_\text{eff} = 0.1$.}
\label{fig:darkradiation}
\end{figure*}

As discussed in \Sec{formalism}, the signal in a direct deflection experiment depends strongly on the relative motion between the plasma and laboratory frames. For cosmological sources of millicharged radiation, the preferred frames of the laboratory and plasma approximately coincide, with a small offset controlled by the velocity $\sim 10^{-3}$ of the solar system. To leading order, we can ignore this correction and adopt $\g = 1$ in quantifying the relative boost between the plasma and laboratory frames. In regards to the particular experimental setup, the use of RF cavities employing $\wdef \, \Ldef \sim 1$ is crucial to generating a signal in this case, since it vanishes in the quasistatic limit ($\wdef \, \Ldef \ll 1$) for $\g = 1$ (see \Sec{formalism}).

The projected sensitivity of Dark SRF to cosmological sources of millicharged dark radiation is shown in \Fig{darkradiation}, fixing the experimental benchmarks as in \Secs{genreach}{darksolarwind}. In the left-panel, we consider two setups of different size, $\Ldef = 10 \ \cm$ and $\Ldef = 1 \ \text{m}$, and show the sensitivity in the $q_\x - \alpha^\p$ plane, fixing the millicharged dark radiation density to $\ODR = 3 \times 10^{-2}$ (corresponding to mCPs produced by dark energy). For the same reasons as discussed in \Sec{darksolarwind}, larger setups can explore smaller millicharges, while smaller cavities are able to probe larger values of the self-coupling (corresponding to stronger plasma backreactions). For small $\alpha^\p$, backreactions from self-interactions of the dark plasma can be ignored. For sufficiently large $q_\x$ and $\alpha^\p$, the dark solar wind from \Sec{darksolarwind} contributes a significant density at Earth, which can potentially modify the local density of an independent cosmological population. Since the projected sensitivity of the experimental setup discussed here does not explore new parameter space in this regime, we do not investigate this possibility further. 

To further explore the weak-coupling regime, in the right-panel of \Fig{darkradiation} we instead show the Dark SRF sensitivity in the $q_\x - \ODR$ plane, fixing $\Ldef = 1 \ \text{m}$ and $\alpha^\p$ to be sufficiently small such that self-interactions can be ignored on laboratory length scales. In this panel, we show the various $\ODR$ benchmarks discussed above, in the case that millicharged dark radiation arises from dark energy, dark matter decay, the cosmic neutrino background, or some other subcomponent of the total primordial radiation density with $\Delta N_\text{eff} = 0.1$. We see that Dark SRF can explore orders of magnitude of new parameter space for a cosmological abundance of millicharged dark radiation, including that arising from the millicharge of the cosmic neutrino background. The sensitivity to such models is parametrically enhanced compared to that of the dark solar wind in \Sec{darksolarwind}. This is largely due to the fact that unlike the dark solar wind, we have considered less-restrictive examples of cosmological dark radiation, treating the density $\ODR$ as a free parameter independent of the couplings $q_\x$ and $\alpha^\p$.

Before concluding this section, we note that terrestrial, solar, and galactic  magnetic fields can significantly impede the propagation of mCPs in the solar system if the interaction mediated by the dark photon is long-ranged on the relevant length scales. For instance, for dark photons longer-ranged than an Earth radius, $\mAp \lesssim R_\oplus^{-1} \sim 10^{-14} \ \eV$, Earth's magnetic field $B_\oplus \sim 0.5 \ \text{G}$ can significantly modify the terrestrial density of such radiation when the mCP gyroradius $r_g \sim \tilde{E}_\x / (e q_\x \, B_\oplus)$ is much smaller than $R_\oplus$~\cite{Emken:2019tni}, where $\tilde{E}_\x \sim 3 \, \tilde{T}_\x$ is the typical energy of mCP radiation. Similarly, for dark photons longer-ranged than a solar radius $\mAp \lesssim R_\odot^{-1} \sim 10^{-16} \ \eV$, we expect solar modulation to suppresses the local abundance of mCPs with energy $\tilde{E}_\x \sim 3 \, \tilde{T}_\x \lesssim 0.2 \ \GeV \times q_\x$~\cite{Cholis:2015gna,Potgieter:2013mcc}. Finally, if the dark photon is long-ranged on galactic length scales, the local abundance of millicharged radiation may be affected by galactic supernova remnants~\cite{Chuzhoy:2008zy,Li:2020wyl}. However, note that these effects need not apply in our parameter space of interest, since we only require that mCPs possess an effective charge on meter-sized length scales.

%%%%%%%%%%%%%%%%%%%%%%%%%%%%%%%%%%%%%%%%%%%%%%%%%%%%%%%%%%%%
\section{Discussion and Conclusion}
\label{sec:conclusion}
%%%%%%%%%%%%%%%%%%%%%%%%%%%%%%%%%%%%%%%%%%%%%%%%%%%%%%%%%%%%

In this work, we have explored a new approach to detect a relativistic background of millicharged particles using light-shining-through-wall experiments, where a cavity is driven with strong electromagnetic fields and placed nearby a quiet shielded cavity. Inadvertently, these setups can also operate as direct deflection experiments. Millicharged radiation passing through the driven cavity is deflected, setting up propagating disturbances of feebly-coupled charges and currents that can resonantly excite small signal fields in the detection cavity. 

We have focused on the existing Dark SRF light-shining-through-wall experiment, since it employs high-$Q$ superconducting RF cavities, which enhances both the strength of the driven fields, as well as the resonant sensitivity of the detection cavity. In particular, our estimates show that a future version of Dark SRF can probe orders of magnitude of unexplored parameter space for relativistic millicharges produced by the Sun or cosmologically in the early or late universe. Such a setup has the potential to measure the abundance of the cosmic neutrino background if it possess a small effective electromagnetic charge, or cosmological dark radiation with an energy density four orders of magnitude smaller than that of the cosmic microwave background. 

In future work, it would be interesting to pursue more dedicated experimental approaches. For instance, multiple deflecting cavities can be used (analogous to a LINAC) in order to multiplicatively increase the signal, or larger magnetic field configurations can be employed to focus the millicharge radiation into a smaller experimental area.  Also note that in the weak-coupling regime, the signal strength in a direct deflection setup is controlled by the millicharge plasma frequency, which scales favorably with larger number densities and smaller characteristic energies. While we have focused on thermal populations of millicharged radiation, this implies that direct deflection setups would have enhanced sensitivity to populations of millicharges with high-occupancy in the lowest momentum-modes, such as those created by parametric resonance or tachyonic instabilities. Generalizations of this approach may also be adapted to search for dark radiation coupled to a non-electromagnetic force, such as one mediated by a new spin-coupled boson. However, in this case an experiment needs to contend with strong constraints on the existence of such new light mediators.  

%%%%%%%%%%%%%%%%%%%%%%%%%%%%%%%%%%%%%%%%%%%%%%%%%%%%%%%%%%%%
\acknowledgments
We thank Paddy Fox for helpful discussions. This research was supported by the U.S.~Department of Energy's Office of Science under contract DE–AC02–76SF00515, as well as by Fermilab's Superconducting Quantum Materials and Systems Center (SQMS) under contract number DE-AC02-07CH11359. Fermilab is operated by the Fermi Research Alliance, LLC under Contract DE-AC02-07CH11359 with the U.S.~Department of Energy. This work was completed in part at the Perimeter Institute. Research at Perimeter Institute is supported in part by the Government of Canada through the Department of Innovation, Science and Economic Development Canada and by the Province of Ontario through the Ministry of Colleges and Universities. E.H.T. also acknowledges support by NSF Grant PHY-2310429, Simons Investigator Award No. 824870, and the Gordon and Betty Moore Foundation Grant GBMF7946.
%%%%%%%%%%%%%%%%%%%%%%%%%%%%%%%%%%%%%%%%%%%%%%%%%%%%%%%%%%%%

\appendix
\addappheadtotoc
\renewcommand\thefigure{\thesection.\arabic{figure}}
\setcounter{figure}{0}

\newpage

%%%%%%%%%%%%%%%%%%%%%%%%%%%%%%%%%%%%%%%%%%%%%%%%%%%%%%%%%%%%
\section{Plasma Formalism Details}
%%%%%%%%%%%%%%%%%%%%%%%%%%%%%%%%%%%%%%%%%%%%%%%%%%%%%%%%%%%%

In this appendix, we provide some of the technical details needed to determine the response of a plasma to an external electromagnetic field. 

%%%%%%%%%%%%%%%%%%%%%%%%%%%%%%%%%%%%%%%%%%%%%%%%%%%%%%%%%%%%
\subsection{Isotropic, Ultrarelativistic, Collisionless Plasma}
\label{app:isoplasma}
%%%%%%%%%%%%%%%%%%%%%%%%%%%%%%%%%%%%%%%%%%%%%%%%%%%%%%%%%%%%

\begin{figure}
    \centering
    \includegraphics[width=0.48\linewidth]{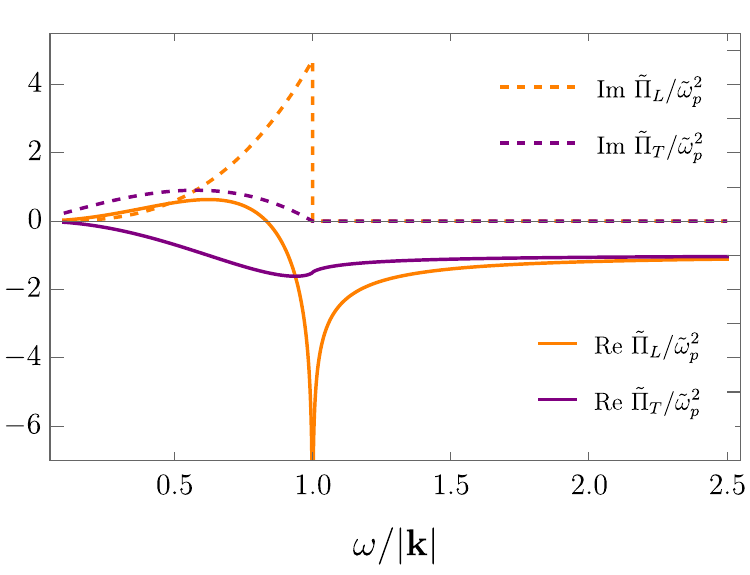}
    \includegraphics[width=0.51\linewidth]{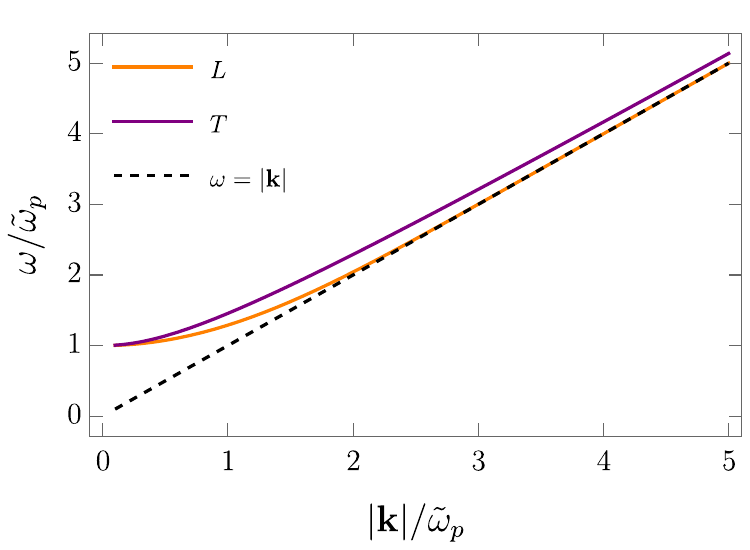}
    \caption{\textbf{Left}: The longitudinal and transverse components of the linear response tensor, given in \Eqs{PiL}{PiT}. \textbf{Right}: The longitudinal ($L$) and transverse ($T$) dispersion relations of an ultrarelativistic plasma, using Eqs.~\ref{eq:dispersionLApp}$-$\ref{eq:PiT}. We also show the light cone $\w=|\kv|$ for comparison. }
    \label{fig:PiLT}
\end{figure}

In the rest frame of the plasma, we take it be isotropic. In this case, the spatial part of the linear response tensor $\tilde{\Pi}^{ij}$ can be decomposed into its longitudinal and transverse components, 
\begin{align}
    \tilde{\Pi}_L\left(\w,|\kv|\right)&=-\frac{k_ik_j}{|\kv|^2} \, \tilde{\Pi}^{ij}(k)\label{eq: PiLdef}\\
    \tilde{\Pi}_T\left(\w,|\kv|\right)&=\frac{1}{2}\left(\eta_{ij}+\frac{k_i k_j}{|\kv|^2}\right) \, \tilde{\Pi}^{ij}(k)\label{eq:PiTdef}
    ~,
\end{align}
as follows~\cite{melrose2008quantum}
\begin{align}
    \tilde{\Pi}^{ij}(k)&=-\frac{k^ik^j}{|\kv|^2} \, \tilde{\Pi}_L\left(\w,|\kv|\right)+\left(\eta^{ij}+\frac{k^ik^j}{|\kv|^2}\right) \, \tilde{\Pi}_T \left(\w,|\kv|\right) 
    ~. \label{eq:LTdecomposition}
\end{align}
Charge continuity and gauge invariance of the induced current $\tilde{J}_\x^\mu(k)=\tilde{\Pi}^{\mu\nu}(k) \, \tilde{A}_\nu(k)$ imply $k_\mu \, \tilde{\Pi}^{\mu\nu}(k)=0$ and $k_\nu \, \tilde{\Pi}^{\mu\nu}(k)=0$, respectively. These can be used to construct the remaining components of $\tilde{\Pi}^{\mu\nu}(k)$ in terms of $\tilde{\Pi}_{L}$~\cite{melrose2008quantum},
\begin{align}
    \tilde{\Pi}^{00}(k)&=\frac{k_ik_j}{\w^2}\tilde{\Pi}^{ij}(k)=-\frac{|\kv|^2}{\w^2}\tilde{\Pi}_L\left(\w,|\kv|\right) \label{eq:Pi00}\\
    \tilde{\Pi}^{i0}(k)&=-\frac{k_j}{\w}\tilde{\Pi}^{ij}(k)=-\frac{k^i}{\w}\tilde{\Pi}_L \left(\w,|\kv|\right)\label{eq:Pii0}\\
    \tilde{\Pi}^{0j}(k)&=-\frac{k_i}{\w}\tilde{\Pi}^{ij}(k)=-\frac{k^j}{\w}\tilde{\Pi}_L\left(\w,|\kv|\right) \label{eq:Pi0i}
    ~.
\end{align}
By requiring that non-trivial solutions $\tilde{A}^\mu(k)\neq 0$ to the sourceless Maxwell's equations exist, we obtain the dispersion relations, which for an \emph{isotropic} plasma decompose into two parts~\cite{melrose2008quantum}
\begin{align}
    \text{Longitudinal:}& &&\w^2+\tilde{\Pi}_L\left(\w,|\kv|\right)=0\label{eq:dispersionLApp}\\
    \text{Transverse:}& &&\w^2-|\kv|^2+\tilde{\Pi}_T\left(\w,|\kv|\right)=0 
    ~. \label{eq:dispersionTApp}
\end{align}
Note that $\tilde{\Pi}_{L,T}\left(\w,|\kv|\right)$ and the corresponding dispersion relations are gauge-invariant~\cite{melrose2008quantum,Ruppert:2005uz}. For an isotropic, ultrarelativistic, and collisionless plasma, the longitudinal and transversal components of the linear response tensor are~\cite{Thoma:2008my,Thoma:2008tb,Chakraborty:2006md,Ruppert:2005uz,Mrowczynski:2007hb}
\begin{align}
    \tilde{\Pi}_L\left(\w,|\kv|\right) &=\left(-\frac{\w^2}{|\kv^2|}\right)\times 3\wpt^2\left(\frac{\w}{2|\kv|}\Lambda-1\right) \label{eq:PiL}\\
    \tilde{\Pi}_T\left(\w,|\kv|\right) &=\left(-1\right)\times \frac{3}{2}\wpt^2\frac{\w^2}{|\kv|^2}\left[1-\left(1-\frac{|\kv|^2}{\w^2}\right)\frac{\w}{2|\kv|}\Lambda\right] \label{eq:PiT}\\
    \Lambda&=\ln\frac{\w+|\kv|}{\w-|\kv|}=\ln\left|\frac{\w+|\kv|}{\w-|\kv|}\right|-i\Theta\left(|\kv|^2-\w^2\right)~, \label{eq:log}
\end{align}
where in the second equality of the last line we have picked the sign of $\text{Im} (\Lambda)$ such that a plane wave $\tilde{\phi}(x)\propto e^{-i\w t+i\kv.\xv}\propto e^{-\text{Im }|\kv|x_\parallel}$ (where $x_\parallel=\xv.\mathbf{\hat{k}}$) propagating in the frame of the plasma with a real positive $\w$ and with $\text{Re }|\kv|>0$ will have $\text{Im }|\kv|>0$, meaning that it decays instead of grows, due to Landau damping when $|\kv|^2>\w^2$.\footnote{From \Eq{dispersionTApp}, we see that $\text{Im }|\kv|=\text{Im }\tilde{\Pi}_T/(2\text{Re }|\kv|)$. Thus, $\text{Im }|\kv|>0$ requires $\text{Im }\tilde{\Pi}_T>0$ for $|k|^2>\w^2$, which is indeed the case for the choice of sign of the imaginary part of $\tilde{\Pi}_T$ in \Eq{log}.} In \App{VlasovDerivation}, we provide a derivation of these expressions for $\tilde{\Pi}_L$ and $\tilde{\Pi}_T$. We plot $\tilde{\Pi}_{L,T}$ and the corresponding dispersion relations in \Fig{PiLT}.

We emphasize that the definitions of $\tilde{\Pi}_{L,T}$ vary in the literature. Our definitions follow that of Ref.~\cite{melrose2008quantum} (M08) and differ from that used in, e.g., Refs.~\cite{Thoma:2008my,Thoma:2008tb,Carrington:2003je,Mandal:2012wi,Mrowczynski:2007hb} (T08) which are, again, different from those used in, e.g., Refs.~\cite{Chakraborty:2006md,Ruppert:2005uz} (C06). If these definitions were to lead to the same longitudinal and transverse dispersion relations, the $\tilde{\Pi}_{L,T}$ must be related as $\tilde{\Pi}_L^{\rm M08}=(-\w^2/|\kv|^2)\tilde{\Pi}_L^{\rm T08}=\left(-\w^2|\kv|^2/k^2\right) \tilde{\Pi}_L^{\rm C06}$ and $\tilde{\Pi}_T^{\rm M08}=(-1)\tilde{\Pi}_T^{\rm T08}=(-1)\tilde{\Pi}_T^{\rm C06}$. Note also that sometimes, e.g., in Ref.~\cite{Mandal:2012wi,Chakraborty:2006md}, the $\tilde{\Pi}_{L,T}$ are expressed in terms of the Debye screening mass $\tilde{m}_D^2=3\wpt^2$~\cite{Chakraborty:2006md,Grayson:2022asf} instead of the what we refer to as the plasma frequency, $\wpt$. In our definition, the plasma frequency $\wpt$ is the lowest frequency at which plasma waves can propagate, 
i.e., the value of $\w$ that solves \Eqs{dispersionLApp}{dispersionTApp} in the limit $|\kv|\rightarrow 0$. The plasma frequency of an ultrarelativistic $e^\pm$ plasma at a temperature $\tilde{T}$, for example, is $\wpt=e\tilde{T}/3$~\cite{Thoma:2008my,Thoma:2008tb,Chu:1988wh,Grayson:2022asf,Mrowczynski:2007hb}. Furthermore, while in our notation $k^2=k_\mu k^\mu$ and $|\kv|^2=k^ik^i$, other variations are often used in the literature.

%%%%%%%%%%%%%%%%%%%%%%%%%%%%%%%%%%%%%%%%%%%%%%%%%%%%%%%%%%%%
\subsection{Vlasov Derivation of the Linear Response Tensor of Ultrarelativistic Pair-Plasma}
\label{app:VlasovDerivation}
%%%%%%%%%%%%%%%%%%%%%%%%%%%%%%%%%%%%%%%%%%%%%%%%%%%%%%%%%%%%

The induced currents $J_{\x}^{\mu}(x)$ in a plasma of $\pm e q_\x$ charged plasma particles $\x^\pm$ can be expressed in terms of its distribution functions $f_\pm (x,p)$~\cite{Chu:1988wh,melrose2008quantum}
\begin{align}
    J_{\x}^{\mu}(x)&=2e q_\x\int \frac{d^3p}{(2\pi)^3}\frac{p^\mu}{p \cdot u} \left[f_+(x,p)-f_-(x,p)\right]=2e q_\x\int \frac{d^3p}{(2\pi)^3}\frac{p^\mu}{p \cdot u}  [\delta f_+(x,p)-\delta f_-(x,p)] \label{eq:Jindintermsoff}
    ~,
\end{align}
where the factor of two accounts for spin degrees of freedom, $u^\mu$ is the velocity of the plasma wind, and the 4-momentum $p^\mu$ is on-shell. In the second equality, we defined $\delta f_\pm=f_\pm-f_{\rm eq}$ as the deviation of the distribution function $f_\pm (x,p)$ of the plasma particles from their thermal equilibrium distribution in the absence of a chemical potential $f_{\rm eq}(p)=\left[e^{p \cdot u/\tilde{T}}+1\right]^{-1}$, and we assumed that the equilibrium plasma has $J_{\x}^\mu(x)=0$. For a collisionless plasma, the evolution of the distribution function $f_\pm (x,p)$ is governed by the Vlasov equation,
\begin{align}
    p^\mu\partial_\mu f_{\pm}(x,p)\mp e q_\x F_{\mu\nu}(x)p^\mu\frac{\partial f_{\pm}(x,p)}{\partial p_\nu}=0
    ~.
\end{align}
In the weak-field regime, this can be solved perturbatively in $A^{\mu}$. To first order in $A^\mu$, and after Fourier-transforming, the solution is given by
\begin{align}
    \delta f_{\pm}(k,p)=& \pm \frac{e q_\x}{\tilde{T}} f_{\rm eq}(p)\left[1-f_{\rm eq}(p)\right] \left[u\cdot A(k)-\frac{(k \cdot u)[p\cdot A(k)]}{k \cdot p} \right] \label{eq:deltaf}
    ~,
\end{align}
where $p$ is the 4-momentum of a $\x$ particle, and $k$ is the wavenumber associated with coordinate position $x$. Since this is linear in $A^\mu$, the resulting induced dark current $J_{\x}^{\mu}$ can be written as
\begin{align}
    J_{\x}^{ \mu}(k)=\Pi^{\mu\nu}(k,u) A_{\nu}(k) 
    ~. \label{eq:linearresponseapp}
\end{align}
Matching \Eqs{Jindintermsoff}{deltaf} with \Eq{linearresponseapp}, we find
\begin{align}
    \Pi^{\mu\nu}(k,u)=&\frac{4(e q_\x)^2}{\tilde{T}}\int \frac{d^3p}{(2\pi)^3(p \cdot u)}f_{\rm eq}(p)\left[1-f_{\rm eq}(p)\right] \left[p^\mu u^\nu-\frac{(k \cdot u)p^\mu p^\nu}{(k \cdot p)}\right]
    ~.
\end{align}
Note that the description in this subsection is thus far covariant. To evaluate the above integral, we: (1) move to the plasma frame by setting $u^\mu=\delta_{\mu 0}$, (2) take the ultrarelativistic plasma temperature limit $p^0\simeq |\mathbf{p}|$, (3) define $\hat{p}^\mu=p^\mu/|\mathbf{p}|$, and (4) define $\cos\theta=\hat{p}^ik^i/|\kv|$. The integral then simplifies as follows,
\begin{align}
    \tilde{\Pi}^{\mu\nu}(k)&=\frac{2(e q_\x)^2}{\tilde{T}}\int_0^\infty \frac{|\mathbf{p}|^2d|\mathbf{p}|}{\pi^2}f_{\rm eq}(|\mathbf{p}|)\left[1-f_{\rm eq}(|\mathbf{p}|)\right]\hat{\Pi}^{\mu\nu}=3\wpt^2\hat{\Pi}^{\mu\nu} 
    ~, \label{eq:PihatPi}
\end{align}
where we have expressed the $|\mathbf{p}|$ integral in terms of the plasma frequency $\wpt=e q_\x \tilde{T}/3$,
\begin{align}
    \frac{2(e q_\x)^2}{\tilde{T}}\int_0^\infty  \frac{|\mathbf{p}|^2d|\mathbf{p}|}{\pi^2}f_{\rm eq}(|\mathbf{p}|)\left[1-f_{\rm eq}(|\mathbf{p}|)\right]=3\wpt^2
    ~,
\end{align}
and collected the remaining angular integral in
\begin{align}
    \hat{\Pi}^{\mu\nu}=\int \frac{d\Omega}{4\pi}\left[\hat{p}^\mu\delta_{\nu 0}-\frac{\w}{\w-|\kv|\cos\theta}\hat{p}^\mu\hat{p}^\nu\right] 
    ~. \label{eq:hatPi}
\end{align}
Using Eqs.~\ref{eq: PiLdef}, \ref{eq:PihatPi}, and \ref{eq:hatPi}, the longitudinal part of $\tilde{\Pi}^{\mu\nu}(k)$ can be evaluated as
\begin{align}
    \tilde{\Pi}_L\left(\w,|\kv|\right)&=-\frac{k_i k_j}{|\kv|^2}\left(3\wpt^2\hat{\Pi}^{ij}\right)=3\wpt^2\int \frac{d\Omega}{4\pi}\frac{\cos^2\theta}{1-(|\kv|/\w)\cos\theta}=-3\wpt^2\frac{\w^2}{|\kv|^2} \left[\frac{\w}{2|\kv|}\Lambda-1\right]
    ~.
\end{align}
Using Eqs.~\ref{eq:PiTdef}, \ref{eq:PihatPi},  and \ref{eq:hatPi}, the transverse part of $\tilde{\Pi}^{\mu\nu}$ can be evaluated as
\begin{align}
    \tilde{\Pi}_T\left(\w,|\kv|\right)&=\frac{1}{2}\left(\eta_{ij}+\frac{k_i k_j}{|\kv|^2}\right)\left(3\w_p^2\hat{\Pi}^{ij}\right)=3\wpt^2\int \frac{d\Omega}{4\pi}\frac{1}{2}\frac{1-\cos^2\theta}{1-(|\kv|/\w)\cos\theta}
    \nl
    &=-3\wpt^2\frac{\w^2}{2|\kv|^2}\left[1-\left(1-\frac{|\kv|^2}{\w^2}\right)\frac{\w}{2|\kv|}\Lambda\right]
    ~,
\end{align}
in agreement with Eqs~\ref{eq:PiL}$-$\ref{eq:log}, which were adapted from Refs.~\cite{Thoma:2008my,Thoma:2008tb,Chakraborty:2006md,Ruppert:2005uz,Mrowczynski:2007hb}. $\tilde{\Pi}_{L,T}$ specifies all of the entries of the linear response tensor in the plasma frame $\tilde{\Pi}^{\mu\nu}(k)$ via Eqs.~\ref{eq:LTdecomposition}$-$\ref{eq:Pi0i}. $\Pi^{\mu\nu}(k)$ in an arbitrary frame can be found by the substitution $\omega\rightarrow k \cdot u$ and $\omega^2-|\mathbf{k}|^2\rightarrow k^2$.

%%%%%%%%%%%%%%%%%%%%%%%%%%%%%%%%%%%%%%%%%%%%%%%%%%%%%%%%%%%%
\subsection{Solving Maxwell's Equations in Coulomb Gauge}
\label{app:MaxwellCoulombgauge}
%%%%%%%%%%%%%%%%%%%%%%%%%%%%%%%%%%%%%%%%%%%%%%%%%%%%%%%%%%%%

In order to invert the Fourier-transformed Maxwell's equations in the plasma frame (\Eq{MaxwellEq}), one needs to pick a gauge. In this paper, we adopt  Coulomb gauge 
\begin{align}
    k_i\tilde{A}^i(k)=0 \label{eq:Coulomb}
    ~.
\end{align}
Maxwell's equations then reduce to
    \begin{align}
    \left(|\kv|^2-\tilde{\Pi}^{00}\right)\tilde{A}_{0}(k)-\tilde{\Pi}^{0i}\tilde{A}_{i}(k)&=\tilde{J}_{\rm def}^0(k)\label{eq:darkMaxwell0}\\
    \left(k^i\w-\tilde{\Pi}^{i0}\right)\tilde{A}_{0}(k)-\left(k^2\eta^{ij}+\tilde{\Pi}^{ij}\right)\tilde{A}_{j}(k)&=\tilde{J}_{\rm def}^i(k) ~. \label{eq:darkMaxwelli}
    \end{align}
Moreover, \Eqs{Coulomb}{LTdecomposition} imply
\begin{align}
    \tilde{\Pi}^{ij}\tilde{A}_{j}(k)&=\tilde{\Pi}_T \tilde{A}^i(k) 
    ~. \label{eq:Piij}
\end{align}
Using Eqs.~\ref{eq:Pi00}, \ref{eq:Pi0i}, and \ref{eq:Coulomb} in \Eq{darkMaxwell0} and Eqs.~\ref{eq:Pii0}, \ref{eq:Piij}, and \ref{eq:Coulomb} in \Eq{darkMaxwelli}, we find
\begin{align}
    \tilde{A}^0(k)&=\frac{1}{|\kv|^2}\frac{\tilde{J}_{\rm def}^0(k)}{1+\tilde{\Pi}_L/\w^2}\\
    \tilde{A}^i(k)&=\frac{\tilde{J}_{\rm def}^i(k)-(k^i\w/|\kv|^2)\tilde{J}_{\rm def}^0(k)}{|\kv|^2-\w^2-\tilde{\Pi}_T}
    ~.
\end{align}
The induced currents are given by \Eq{LinResp1}
\begin{align}
    \tilde{J}_{\x}^0(k)=&-\frac{\tilde{\Pi}_L}{\w^2+\tilde{\Pi}_L}\tilde{J}_{\rm def}^0(k) \label{eq:rhoind}\\
    \tilde{J}_{\x}^i(k)=&\frac{\tilde{\Pi}_T}{|\kv|^2-\w^2-\tilde{\Pi}_T}\tilde{J}_{\rm def}^i(k)-\left(\frac{\tilde{\Pi}_L}{\w^2+\tilde{\Pi}_L}+\frac{\tilde{\Pi}_T}{|\kv|^2-\w^2-\tilde{\Pi}_T}\right)\frac{k^i\w}{|\kv|^2}\tilde{J}_{\rm def}^0(k) ~, \label{eq:Jiind}
\end{align}
where we have used Eqs.~\ref{eq:Pi00}, \ref{eq:Pi0i}, and \ref{eq:Coulomb} to arrive at the first line and used Eqs.~\ref{eq:Pii0}, \ref{eq:Piij}, and \ref{eq:Coulomb} to arrive at the second line. It can be shown that if the external currents obey charge continuity, $k_\mu\tilde{J}_{\rm def}^\mu(k)=0$, then the induced currents also obey charge continuity, $k_\mu\tilde{J}_{\x}^\mu(k)=0$, manifestly.

%%%%%%%%%%%%%%%%%%%%%%%%%%%%%%%%%%%%%%%%%%%%%%%%%%%%%%%%%%%%
\section{Response of an Ultrarelativistic Plasma to an Oscillating Deflector}
\label{app:acdef}
%%%%%%%%%%%%%%%%%%%%%%%%%%%%%%%%%%%%%%%%%%%%%%%%%%%%%%%%%%%%
\label{app:inverseFourier}

%%%%%%%%%%%%%%%%%%%%%%%%%%%%%%%%%%%%%%%%%%%%%%%%%%%%%%%%%%%%
\subsection{Zero Wind Velocity}
%%%%%%%%%%%%%%%%%%%%%%%%%%%%%%%%%%%%%%%%%%%%%%%%%%%%%%%%%%%%
Here, we derive the expressions for the induced currents $J_\chi^\mu$ in the absence of a plasma wind in the laboratory frame, $\gamma=1$. The Fourier-transformed deflector currents are
\begin{align}
    J_{\rm def}^{0}(k)=& \left(-k^z\right)i4\pi^{7/2}\mathcal{J}_{\rm def}\Ldef^2e^{-\frac{\left(k^z\right)^2\Ldef^2}{4}}\Delta_{\gamma=1}(k) \label{eq:Jdef0kzerowind}\\
    J_{\rm def}^{z}(k)=&\left(-\wdef\right)i4\pi^{7/2}\mathcal{J}_{\rm def}\Ldef^2 e^{-\frac{\left(k^z\right)^2\Ldef^2}{4}}\Delta_{\gamma=1}(k) ~, \label{eq:Jdefzkzerowind}
\end{align}
where 
\begin{align}
    \Delta_{\gamma=1}(k)=&\delta\left(\w-\wdef\right) \left[\frac{\delta\left(k^x-\Ldef^{-1}\right)+\delta\left(k^x+\Ldef^{-1}\right)}{2}\right]\left[\frac{\delta\left(k^y-\Ldef^{-1}\right)+\delta\left(k^y+\Ldef^{-1}\right)}{2}\right]~. \label{eq:Diracdeltas2}
\end{align}
From Eqs.~\ref{eq:rhoind}, \ref{eq:Jiind}, \ref{eq:Jdef0kzerowind}, and \ref{eq:Jdefzkzerowind}, we find
\begin{align}
    J_\x^{0,z}(x)&=\frac{i\mathcal{J}_{\rm def}\Ldef^2}{4\pi^{1/2}}e^{-i\wdef t}\cos\left(\frac{x}{\Ldef}\right)\cos\left(\frac{y}{\Ldef}\right)\int_{-\infty}^{+\infty}dk^z e^{ik^z z}e^{-\frac{\left(k^z\right)^2\Ldef^2}{4}}\left[\Xi_{\gamma=1}^{0,z}(k)\right]_{\w=\wdef,\,|k^{x,y}|=\Ldef^{-1}}
    \label{eq:nowindjchi1}\\
    J_\x^x(x)&=-\frac{\mathcal{J}_{\rm def}\Ldef^2}{4\pi^{1/2}}e^{-i\wdef t}\sin\left(\frac{x}{\Ldef}\right)\cos\left(\frac{y}{\Ldef}\right)\int_{-\infty}^{+\infty}dk^z e^{ik^z z}e^{-\frac{\left(k^z\right)^2\Ldef^2}{4}}\left[\Xi_{\gamma=1}^{x}(k)\right]_{\w=\wdef,\,|k^{x,y}|=\Ldef^{-1}} \label{eq:nowindjchi2}\\
    J_\x^y(x)&=-\frac{\mathcal{J}_{\rm def}\Ldef^2}{4\pi^{1/2}}e^{-i\wdef t}\cos\left(\frac{x}{\Ldef}\right)\sin\left(\frac{y}{\Ldef}\right)\int_{-\infty}^{+\infty}dk^z e^{ik^z z}e^{-\frac{\left(k^z\right)^2\Ldef^2}{4}}\left[\Xi_{\gamma=1}^{y}(k)\right]_{\w=\wdef,\,|k^{x,y}|=\Ldef^{-1}}
    ~,
    \label{eq:nowindjchi3}
\end{align}
where the remaining $k^z$ integral must be done numerically, and
\begin{align}
    \Xi_{\gamma=1}^\mu(k)=
    \begin{cases}
       \displaystyle -\frac{\Pi_L}{\w^2+\Pi_L}\left(-k^z\right) &\text{ for }\mu=0\\
    \displaystyle\frac{\Pi_T}{|\kv|^2-\w^2-\Pi_T}\left(-\wdef\right)-\left(\frac{\Pi_L}{\w^2+\Pi_L}+\frac{\Pi_T}{|\kv|^2-\w^2-\Pi_T}\right)\frac{k^z\w}{|\kv|^2}\left(-k^z\right) &\text{ for }\mu=z\\
    \displaystyle -\left(\frac{\Pi_L}{\w^2+\Pi_L}+\frac{\Pi_T}{|\kv|^2-\w^2-\Pi_T}\right)\frac{\Ldef^{-1}\,\w}{|\kv|^2}\left(-k^z\right) &\text{ for }\mu=x,y~.
    \end{cases}
\end{align}

%%%%%%%%%%%%%%%%%%%%%%%%%%%%%%%%%%%%%%%%%%%%%%%%%%%%%%%%%%%%
\subsection{Non-zero Wind Velocity}
%%%%%%%%%%%%%%%%%%%%%%%%%%%%%%%%%%%%%%%%%%%%%%%%%%%%%%%%%%%%

Above, we have adopted the notation where the presence (absence) of a tilde on a function indicates that the function \textit{including its argument} is evaluated in the plasma (laboratory) frame. In this subsection only, to keep expressions concise we abuse this notation by sometimes writing lab-frame quantities as functions of plasma-frame coordinates, e.g., $J_{\rm def}^0(\tilde{k})$, or vice versa, e.g., $\tilde{J}_\x^0(x)$. 

In the case of a non-zero plasma wind, $\gamma>1$, we first compute quantities of interest in the plasma frame, where the plasma response is easier to understand, and then Lorentz transform to the lab frame. We start by expressing the \textit{lab-frame} deflector currents $J_{\rm def}^{\mu}(x)$ in terms of \textit{plasma-frame} coordinates $J_{\rm def}^{\mu}(\tilde{x})$ using the following Lorentz tranformation of coordinates 
\begin{align}
    \tilde{t}=\g\left(t+v z\right),\quad 
    \tilde{z}=\g\left(z+v t\right),\quad 
    \tilde{x}=x,\quad 
    \tilde{y}=y ~. \label{eq:Lorentzcoord}    
\end{align}
Next, we Fourier transform $J_{\rm def}^{\mu}(\tilde{x})$ with respect to the \textit{plasma-frame} coordinates $\tilde{x}$, 
\begin{align}
    J_{\rm def}^{0}\left(\tilde{k}\right)=&\left(\g\wdef-\tilde{\w}\right)\frac{i4\pi^{7/2}\mathcal{J}_{\rm def}\Ldef^2}{\g^2 v}e^{-\frac{\left(\tilde{\w}-\g\wdef\right)^2\Ldef^2}{4\g^2v^2}}\Delta_{\g>1}\left(\tilde{k}\right)\\
     J_{\rm def}^{z}\left(\tilde{k}\right)=&\left(-\g\wdef v\right)\frac{i4\pi^{7/2}\mathcal{J}_{\rm def}\Ldef^2}{\g^2 v}e^{-\frac{\left(\tilde{\w}-\g\wdef\right)^2\Ldef^2}{4\g^2v^2}}\Delta_{\g>1}\left(\tilde{k}\right)
     ~,
\end{align}
where 
\begin{align}
    \Delta_{\g>1}\left(\tilde{k}\right)=&\delta\left(\tilde{\w}-\frac{\wdef}{\g}-\tilde{k}^zv\right) \left[\frac{\delta\left(k^x-\Ldef^{-1}\right)+\delta\left(k^x+\Ldef^{-1}\right)}{2}\right]\left[\frac{\delta\left(k^y-\Ldef^{-1}\right)+\delta\left(k^y+\Ldef^{-1}\right)}{2}\right] ~.\label{eq:Diracdeltas}
\end{align}
Using the Lorentz transformations, 
\begin{align}
    \tilde{J}_{\rm def}^0=\g\left(J_{\rm def}^{0}+vJ_{\rm def}^{z}\right),\quad \tilde{J}_{\rm def}^z=\g\left(J_{\rm def}^{z}+vJ_{\rm def}^{0}\right),\quad 
    \tilde{J}_{\rm def}^x=J_{\rm def}^{x},\quad 
    \tilde{J}_{\rm def}^y=J_{\rm def}^{y}~, \label{eq:Lorentzcurrent}
\end{align}
we obtain the momentum-space deflector currents in the plasma frame,
\begin{align}
        \tilde{J}_{\rm def}^{0}\left(\tilde{k}\right)=&\left(\frac{\wdef}{\g}-\tilde{\w}\right)\frac{i4\pi^{7/2}\mathcal{J}_{\rm def}\Ldef^2}{\g v}e^{-\frac{\left(\tilde{\w}-\g\wdef\right)^2\Ldef^2}{4\g^2v^2}}\Delta_{\g>1}\left(\tilde{k}\right)\label{eq:rhoextFourier}\\
        \tilde{J}_{\rm def}^{z}\left(\tilde{k}\right)=&\left(-\tilde{\w} v\right)\frac{i4\pi^{7/2}\mathcal{J}_{\rm def}\Ldef^2}{\g v}e^{-\frac{\left(\tilde{\w}-\g\wdef\right)^2\Ldef^2}{4\g^2v^2}}\Delta_{\g>1}\left(\tilde{k}\right)~.\label{eq:JzextFourier}
\end{align}
Then, Eqs.~\ref{eq:rhoind} and \ref{eq:rhoextFourier} as well as Eqs.~\ref{eq:Jiind}, \ref{eq:rhoextFourier}, \ref{eq:JzextFourier}, and \ref{eq:Lorentzcoord} give\footnote{Here, we use the three delta functions in \Eq{Diracdeltas} to evaluate the $k^x, k^y, \tilde{k}^z$ integrals, leaving the $\tilde{\omega}$ integral as the remaining non-trivial integral. Since one of the Dirac deltas imposes $\tilde{k}^z=(\tilde{\omega}-\omega_{\rm def}/\gamma)/v$, one cannot easily take the $v\rightarrow 0$ limit to recover the zero wind velocity results. Nevertheless, we checked numerically that the $\gamma>1$ results reduce to the $\gamma=1$ ones as we bring $\gamma$ close to 1. Had we evaluated the $\tilde{\omega}, k^x, k^y$ first using the three delta functions, we would be left with a $\tilde{k}^z$ integral instead. In the limit $v\rightarrow 0$, this $\tilde{k}^z$ integral reduces trivially to the $\gamma=1$ results, Eqs. \ref{eq:nowindjchi1}$-$\ref{eq:nowindjchi3}.}
\begin{align}
        \tilde{J}_\x^{0,z}(x)&=\frac{i\mathcal{J}_{\rm def}\Ldef^{2}}{4\pi^{1/2}\g v^2}e^{-i\wdef t}\cos\left(\frac{x}{\Ldef}\right)\cos\left(\frac{y}{\Ldef}\right)\int_{-\infty}^\infty d\tilde{\w}\, e^{i\frac{\tilde{\omega}-\gamma\omega_{\rm def}}{\gamma v}z}e^{-\frac{\left(\tilde{\w}-\g\wdef\right)^2\Ldef^2}{4\g^2 v^2}}\left[\Xi_{\g>1}^{0,z}\left(\tilde{k}\right)\right]_{\tilde{k}^z=\frac{\tilde{\w}-\wdef/\g}{v},|k^{x,y}|=\Ldef^{-1}} 
    \nl
    \tilde{J}_\x^x(x)&=-\frac{\mathcal{J}_{\rm def}\Ldef^2}{4\pi^{1/2}\gamma v^2}e^{-i\wdef t}\sin\left(\frac{x}{\Ldef}\right)\cos\left(\frac{y}{\Ldef}\right)\int_{-\infty}^\infty d\tilde{\w}\, e^{i\frac{\tilde{\omega}-\gamma\omega_{\rm def}}{\gamma v}z}e^{-\frac{\left(\tilde{\w}-\g\wdef\right)^2\Ldef^2}{4\g^2 v^2}}\left[\Xi_{\g>1}^{x}\left(\tilde{k}\right)\right]_{\tilde{k}^z=\frac{\tilde{\w}-\wdef/\g}{v},|k^{x,y}|=\Ldef^{-1}} 
    \nl
    \tilde{J}_\x^y(x)&=-\frac{\mathcal{J}_{\rm def}\Ldef^2}{4\pi^{1/2}\gamma v^2}e^{-i\wdef t}\cos\left(\frac{x}{\Ldef}\right)\sin\left(\frac{y}{\Ldef}\right)\int_{-\infty}^\infty d\tilde{\w}\, e^{i\frac{\tilde{\omega}-\gamma\omega_{\rm def}}{\gamma v}z}e^{-\frac{\left(\tilde{\w}-\g\wdef\right)^2\Ldef^2}{4\g^2 v^2}}\left[\Xi_{\g>1}^{y}\left(\tilde{k}\right)\right]_{\tilde{k}^z=\frac{\tilde{\w}-\wdef/\g}{v},|k^{x,y}|=\Ldef^{-1}} 
\end{align}
where
\begin{align}
    \Xi_{\g>1}^\mu\left(\tilde{k}\right)=
    \begin{cases}
       \displaystyle -\frac{\tilde{\Pi}_L}{\tilde{\w}^2+\tilde{\Pi}_L}\left(\frac{\wdef}{\g}-\tilde{\w}\right) &\text{ for }\mu=0\\
        \displaystyle\frac{\tilde{\Pi}_T}{|\tilde{\kv}|^2-\tilde{\w}^2-\tilde{\Pi}_T}\left(-\tilde{\w} v\right)-\left(\frac{\tilde{\Pi}_L}{\tilde{\w}^2+\tilde{\Pi}_L}+\frac{\tilde{\Pi}_T}{|\tilde{\kv}|^2-\tilde{\w}^2-\tilde{\Pi}_T}\right)\frac{\tilde{k}^z\tilde{\w}}{|\tilde{\kv}|^2}\left(\frac{\wdef}{\g}-\tilde{\w}\right) &\text{ for }\mu=z\\
        \displaystyle -\left(\frac{\tilde{\Pi}_L}{\tilde{\w}^2+\tilde{\Pi}_L}+\frac{\tilde{\Pi}_T}{|\tilde{\kv}|^2-\tilde{\w}^2-\tilde{\Pi}_T}\right)\frac{\Ldef^{-1}\,\tilde{\w}}{|\tilde{\kv}|^2}\left(\frac{\wdef}{\g}-\tilde{\w}\right) &\text{ for }\mu=x,y~.
    \end{cases}
\end{align}
Finally, the lab-frame induced currents $J_\x^\mu(x)$ are found using the inverse of \Eq{Lorentzcurrent}.

%%%%%%%%%%%%%%%%%%%%%%%%%%%%%%%%%%%%%%%%%%%%%%%%%%%%%%%%%%%%
\subsection{Regimes of Plasma Response}
%%%%%%%%%%%%%%%%%%%%%%%%%%%%%%%%%%%%%%%%%%%%%%%%%%%%%%%%%%%%

It is well known that an oscillating deflector at rest in the plasma frame either excites on-shell, propagating plasma waves or gets Debye shielded, depending on whether its frequency is above or below the plasma frequency $\wpt$~\cite{Weldon:1982aq, Ahonen:1998iz}. The boundary between the two regimes becomes less trivial when the deflector is not at rest in the plasma frame. In this section, we chart the different regimes of the ultrarelativistic-plasma response to a moving deflector. 

%%%%%%%%%%%%%%%%%%%%%%%%%%%%%%%%%%%%%%%%%%%%%%%%%%%%%%%%%%%%
\subsubsection{Weak backreaction: $\wpt\ll \Ldef^{-1}$}
%%%%%%%%%%%%%%%%%%%%%%%%%%%%%%%%%%%%%%%%%%%%%%%%%%%%%%%%%%%%

The longitudinal and transverse propagators in the plasma can be inferred from \Eq{A0Ai}. They reduce to those of the vacuum when $\tilde{\w}^2\gg |\tilde{\Pi}_L|$ in the longitudinal case and $k_\mu k^\mu\gg |\tilde{\Pi}_T|$ in the transverse case. The deflector can excite modes with $\w\lesssim \wdef\sim \Ldef^{-1}$, $k_{x,y}=L_{\rm def}^{-1}$, and $|k^z|\lesssim \Ldef^{-1}$, which means typically $\tilde{\w}=\g\left(\w-vk^z\right)\lesssim \g\Ldef^{-1}$ and $k_\mu k^\mu\sim \Ldef^{-2}$.
On the other hand, as displayed in \Fig{PiLT}, we have $|\tilde{\Pi}_L|\sim |\tilde{\Pi}_T|\lesssim \wpt^2$.\footnote{This is true unless $\tilde{\w}/|\tilde{\kv}|\simeq 1$ which has a negligible phase-space measure in the weak-backreaction regime, but more generally can be important as discussed below.} Hence, the longitudinal and transverse plasma responses are negligible when $\tilde{\omega}_p\ll \gamma\Ldef^{-1}$ and $\wpt\ll \Ldef^{-1}$, respectively. We refer to the regime $\tilde{\omega}_p\ll  \Ldef^{-1}$, where both the longitudinal and transverse responses are negligible, as the \textit{weak-backreaction} regime. Intuitively, in this regime the deflector operates at such a high plasma-frame frequency/wavenumber that the plasma does not have enough time to react in a significant way. Using \Eq{LinResp1} in the lab frame and using that the magnitude of the linear response tensor is typically $|\Pi^{\mu\nu}|\sim \wpt^2$, we can estimate the magnitude of the induced current as $J_\x\sim  \wpt^2 \, \phi$, where $\phi$ is the electric potential. Since the plasma response is negligible in this regime, the total electric potential is approximately equal to that sourced by the deflector, $\phi\simeq \phi_{\rm def}\sim B_{\rm def} \Ldef$. It follows that
\begin{align}
    J_{\x}\sim \left(\wpt \, \Ldef\right)^2\phi_{\rm def}
    ~.
\end{align}

Outside of the weak-backreaction regime, the gauge field $A_\x^\mu$ sourced by the induced plasma current $J_\x^\mu$ contributes significantly to the total gauge field $A^\mu$ which, in turn, affects the generation of $J_\x^\mu$. In that case, the gauge field $A^\mu$ must be solved self-consistently to all orders in the plasma frequency $\wpt$.

%%%%%%%%%%%%%%%%%%%%%%%%%%%%%%%%%%%%%%%%%%%%%%%%%%%%%%%%%%%%
\subsubsection{On-shell plasmon excitations: $\tilde{\omega}_p\lesssim \gamma \Ldef^{-1}$}
%%%%%%%%%%%%%%%%%%%%%%%%%%%%%%%%%%%%%%%%%%%%%%%%%%%%%%%%%%%%

\begin{figure}
    \centering
    \includegraphics[width=0.505\linewidth]{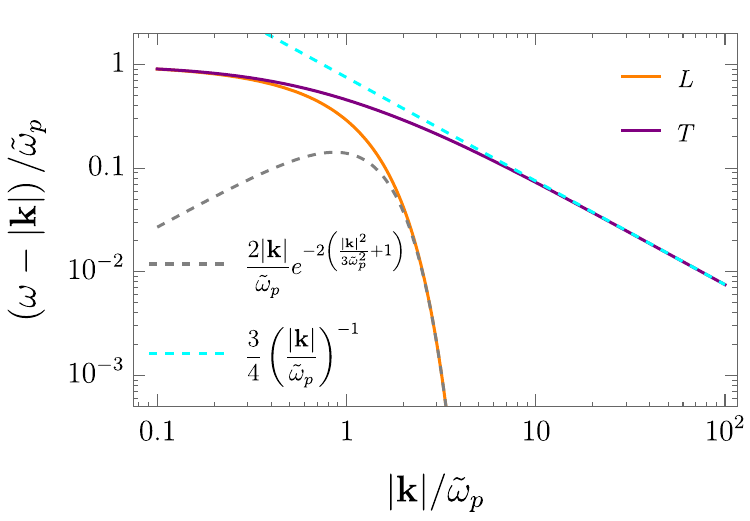}
    \caption{The deviation of the dispersion relation from the light cone $\w=|\kv|$. Analytical approximations to the dispersion relations, which are valid at $|\kv|/\w_p\gtrsim 2$, are also shown.}
    \label{fig:dispersionexpandpower}
\end{figure}

The Fourier-transformed deflector current in the plasma frame has the form $\tilde{J}_{\rm def}^\mu(k)\propto \delta \left[\wdef/\g-k_\mu v^\mu\right]$, where $v^\mu$ is the four-velocity of the deflector. Thus, only modes satisfying $k_\mu v^\mu=\wdef/\gamma$ can be excited by the deflector.\footnote{Alternatively,  this resonance condition can also be seen in position-space where the deflector current reads $\tilde{J}_{\rm def}^\mu(x)\propto e^{-i\Phi_{\rm def}}$, with $\Phi_{\rm def}=\wdef \, t=\g \, \wdef \, \left(\tilde{t}-v\tilde{z}\right)$ as per a simple Lorentz transformation $t=\gamma(\tilde{t}-v\tilde{z})$. Along the worldline of a particle moving with the deflector $\tilde{z}=v\tilde{t}+\text{constant}$, the deflector phase evolves as $d\Phi_{\rm def}/d\tilde{t}=\wdef/\g$ (i.e., the usual special-relativistic time-dilation effect), which is to be matched with the Doppler shifted frequencies of plasma waves $k_\mu v^\mu$. Note also that the Cherenkov resonance condition for a static deflector, $k_\mu v^\mu=\w-\kv.\mathbf{v}=0$, is recovered in the limit $\wdef\rightarrow 0$~\cite{melrose2008quantum,lifschitz1983physical}.} Plasma modes satisfying this resonance condition are excited by the deflector. However, unless they are on-shell, they will subsequently decay due to Landau damping. Propagating (on-shell) plasma modes can be excited if the $(\w, \kv)$ phase-space defined by $k_\mu v^\mu=\wdef/\gamma$ intersects with the plasma dispersion relations, displayed in \Figs{PiLT}{dispersionexpandpower}. To see if this is the case, we rewrite $k_\mu v^\mu=\wdef/\gamma$ in terms of $\cos\tilde{\theta}=\tilde{k}_z/|\tilde{\kv}|$,
\begin{align}
    \frac{|\tilde{\kv}|}{\wpt}\left(1-v\cos\tilde{\theta}\right)+\frac{\tilde{\w}-|\tilde{\kv}|}{\wpt}\label{eq:freqmatchingcos}=\frac{\wdef}{\g\wpt}~.
\end{align}
Let us determine the smallest $\wdef$ relative to $\g \, \wpt$ that leads to excitation of on-shell plasma waves. This amounts to finding the minimum value of the left-hand side when the dispersion relations are imposed. The first term can be as small as $\sim |\tilde{\kv}|/(\g^2\wpt)$ when $\tilde{\theta}\lesssim 1/\g\ll 1$.\footnote{Excited modes with $\tilde{\theta}\lesssim 1/\g$ in the plasma frame can have lab-frame angles of $\theta=\order{1}$, which follows from inverse relativistic-beaming $\cos\theta=(\cos\tilde{\theta}-v)/(1-v\cos\tilde{\theta})$.} 
The second term, although more involved, can be understood through \Fig{dispersionexpandpower}, which indicates that 
\begin{align}
    \frac{\tilde{\w}_L-|\tilde{\kv}|}{\wpt}\simeq \text{min}\left[\order{1} \, , \, \frac{2|\tilde{\kv}|}{\wpt}e^{-2\left(\frac{|\tilde{\kv}|^2}{3\wpt^2}+1\right)}\right]
    ~~,~~
    \frac{\tilde{\w}_T-|\tilde{\kv}|}{\wpt}\simeq \text{min}\left[\order{1} \, , \, 
       \frac{3\wpt}{4|\tilde{\kv}|}\right]~.
\end{align}
Considering the sum of $\sim |\tilde{\kv}|/(\g^2\wpt)$ and each of the above, we find that the left-hand side of \Eq{freqmatchingcos} is minimized  when $|\tilde{\kv}|/\wpt\sim 2$ at a value $\sim \g^{-2}$ for longitudinal modes, while it is minimized when $|\tilde{\kv}|/\wpt\sim \g$ at a value $\sim \g^{-1}$ for transverse modes.\footnote{Since the first and second terms in the left-hand side of \Eq{freqmatchingcos} are both positive and have opposite monotonic behaviors, their sum is minimized when these terms are comparable.} Therefore, for $\wdef\sim \Ldef^{-1}$, the conditions for exciting on-shell plasma waves are $\wpt\lesssim \g\Ldef^{-1}$ for longitudinal modes and $\wpt\lesssim\Ldef^{-1}$ for transverse modes. In cases where the plasma frequency lies in the range $\Ldef^{-1}\lesssim \tilde{\omega_p}\lesssim \g \Ldef^{-1}$, the deflector excites on-shell plasmons which backreact significantly on the total gauge field $A^\mu$.

%%%%%%%%%%%%%%%%%%%%%%%%%%%%%%%%%%%%%%%%%%%%%%%%%%%%%%%%%%%%
\subsubsection{Dynamical Debye screening: $\tilde{\omega}_p\gtrsim \g \Ldef^{-1}$}
%%%%%%%%%%%%%%%%%%%%%%%%%%%%%%%%%%%%%%%%%%%%%%%%%%%%%%%%%%%%

When $\tilde{\w_p}\gtrsim \tilde{\omega}_{\rm def}\sim \g \Ldef^{-1}$, the deflector excites only plasma waves with imaginary wavenumbers. The latter means the plasma waves decay spatially, i.e., the deflector is Debye screened.

%%%%%%%%%%%%%%%%%%%%%%%%%%%%%%%%%%%%%%%%%%%%%%%%%%%%%%%%%%%%
\section{Electromagnetic Fields of the Deflector}
%%%%%%%%%%%%%%%%%%%%%%%%%%%%%%%%%%%%%%%%%%%%%%%%%%%%%%%%%%%%

The gauge potentials sourced by the deflector in the lab frame are given by
\begin{align}
    A_\text{def}^0(k)=\frac{J_\text{def}^0(k)}{|\mathbf{k}|^2}
    ~~,~~
    \mathbf{A}_\text{def}(k)&=\frac{\mathbf{J}_\text{def}(k)-(\w \mathbf{k}/|\mathbf{k}|^2)J_\text{def}^0}{|\mathbf{k}|^2-\w^2}
    ~.
\end{align}
The deflector's electric and magnetic fields can be computed from these gauge potentials through the following relations
\begin{align}
    \mathbf{E}_\text{def}(k)=-i\mathbf{k}A_\text{def}^0(k)+i\w\mathbf{A}_\text{def}
    ~~,~~ 
    \mathbf{B}_\text{def}(k)=i\mathbf{k}\times \mathbf{A}_\text{def}(k)
    ~.
\end{align}
It follows that
\begin{align}
    E_{\rm def}^z(x)&=\frac{i\mathcal{J}_{\rm def}\Ldef^2}{4\pi^{1/2}}e^{-i\wdef t}\cos\left(\frac{x}{\Ldef}\right)\cos\left(\frac{y}{\Ldef}\right)\int_{-\infty}^{+\infty}dk^z e^{ik^z z}e^{-\frac{\left(k^z\right)^2\Ldef^2}{4}}\left[\Xi_{E}^{z}(k)\right]_{\w=\wdef,\,|k^{x,y}|=\Ldef^{-1}}
    \\
    E_{\rm def}^x(x)&=-\frac{\mathcal{J}_{\rm def}\Ldef^2}{4\pi^{1/2}}e^{-i\wdef t}\sin\left(\frac{x}{\Ldef}\right)\cos\left(\frac{y}{\Ldef}\right)\int_{-\infty}^{+\infty}dk^z e^{ik^z z}e^{-\frac{\left(k^z\right)^2\Ldef^2}{4}}\left[\Xi_{E}^{x}(k)\right]_{\w=\wdef,\,|k^{x,y}|=\Ldef^{-1}}\\
    E_{\rm def}^y(x)&=-\frac{\mathcal{J}_{\rm def}\Ldef^2}{4\pi^{1/2}}e^{-i\wdef t}\cos\left(\frac{x}{\Ldef}\right)\sin\left(\frac{y}{\Ldef}\right)\int_{-\infty}^{+\infty}dk^z e^{ik^z z}e^{-\frac{\left(k^z\right)^2\Ldef^2}{4}}\left[\Xi_{E}^{y}(k)\right]_{\w=\wdef,\,|k^{x,y}|=\Ldef^{-1}}
    ~,
\end{align}
and
\begin{align}
     B_{\rm def}^z(x)&=0
    \\
    B_{\rm def}^x(x)&=\frac{\mathcal{J}_{\rm def}\Ldef^2}{4\pi^{1/2}}e^{-i\wdef t}\cos\left(\frac{x}{\Ldef}\right)\sin\left(\frac{y}{\Ldef}\right)\int_{-\infty}^{+\infty}dk^z e^{ik^z z}e^{-\frac{\left(k^z\right)^2\Ldef^2}{4}}\left[\Xi_{B}(k)\right]_{\w=\wdef,\,|k^{x,y}|=\Ldef^{-1}}\\
    B_{\rm def}^y(x)&=-\frac{\mathcal{J}_{\rm def}\Ldef^2}{4\pi^{1/2}}e^{-i\wdef t}\sin\left(\frac{x}{\Ldef}\right)\cos\left(\frac{y}{\Ldef}\right)\int_{-\infty}^{+\infty}dk^z e^{ik^z z}e^{-\frac{\left(k^z\right)^2\Ldef^2}{4}}\left[\Xi_{B}(k)\right]_{\w=\wdef,\,|k^{x,y}|=\Ldef^{-1}} 
    ~,
\end{align}
where
\begin{align}
    \Xi_E^{i}(k)&=\begin{cases}
        \displaystyle -ik^z\frac{-k^z}{|\mathbf{k}|^2}+i\w\frac{\left(-\w\right)-\left(\w k^z/|\mathbf{k}|^2\right)\left(-k^z\right)}{|\mathbf{k}|^2-\w^2}, &\text{for }i=z\\
        \displaystyle -i\Ldef^{-1}\frac{-k^z}{|\mathbf{k}|^2}+i\w\frac{-\left(\w \Ldef^{-1}/|\mathbf{k}|^2\right)\left(-k^z\right)}{|\mathbf{k}|^2-\w^2}, &\text{for }i=x,y
    \end{cases}\\
    \Xi_B(k)&=\displaystyle -i\Ldef^{-1}\frac{-\omega}{|\mathbf{k}|^2-\w^2}
    ~.
\end{align}
The electric and magnetic fields of the $\text{TM}_{010}$ mode are
\begin{align}
    \mathbf{E}_{\text{TM}_{010}}=-E_0J_0\left(\frac{2.4\sqrt{x^2+y^2}}{R}\right)e^{i\frac{2.4}{R} t}\hat{z}
    ~~,~~
    \mathbf{B}_{\text{TM}_{010}}=iE_0J_1\left(\frac{2.4\sqrt{x^2+y^2}}{R}\right)e^{i\frac{2.4}{R} t}\hat{\phi}
    ~.
\end{align}
where $J_0$ and $J_1$ are Bessel functions of zeroth and first order, respectively.

%%%%%%%%%%%%%%%%%%%%%%%%%%%%%%%%%%%%%%%%%%%%%%%%%%%%%%%%%%%%

\bibliographystyle{utphys3}
\bibliography{references}

\providecommand{\href}[2]{#2}\begingroup\raggedright\begin{thebibliography}{10}

\bibitem{Raffelt:1996wa}
G.~G. Raffelt, {\em {Stars as laboratories for fundamental physics}: {The astrophysics of neutrinos, axions, and other weakly interacting particles}}.
\newblock 5, 1996.

\bibitem{Bogorad:2021uew}
Z.~Bogorad and N.~Toro, ``{Ultralight millicharged dark matter via misalignment},'' \href{https://dx.doi.org/10.1007/JHEP07(2022)035}{{\em JHEP} {\bfseries 07} (2022) 035}, \href{https://arxiv.org/abs/2112.11476}{{\ttfamily arXiv:2112.11476 [hep-ph]}}.

\bibitem{Eby:2024mhd}
J.~Eby and V.~Takhistov, ``{Diffuse Axion Background},'' \href{https://arxiv.org/abs/2402.00100}{{\ttfamily arXiv:2402.00100 [hep-ph]}}.

\bibitem{Nguyen:2023czp}
N.~H. Nguyen, E.~H. Tanin, and M.~Kamionkowski, ``{Spectra of axions emitted from main sequence stars},'' \href{https://dx.doi.org/10.1088/1475-7516/2023/11/091}{{\em JCAP} {\bfseries 11} (2023) 091}, \href{https://arxiv.org/abs/2307.11216}{{\ttfamily arXiv:2307.11216 [hep-ph]}}.

\bibitem{DeRocco:2019jti}
W.~DeRocco, P.~W. Graham, D.~Kasen, G.~Marques-Tavares, and S.~Rajendran, ``{Supernova signals of light dark matter},'' \href{https://dx.doi.org/10.1103/PhysRevD.100.075018}{{\em Phys. Rev. D} {\bfseries 100} no.~7, (2019) 075018}, \href{https://arxiv.org/abs/1905.09284}{{\ttfamily arXiv:1905.09284 [hep-ph]}}.

\bibitem{Hou:2011ec}
Z.~Hou, R.~Keisler, L.~Knox, M.~Millea, and C.~Reichardt, ``{How Massless Neutrinos Affect the Cosmic Microwave Background Damping Tail},'' \href{https://dx.doi.org/10.1103/PhysRevD.87.083008}{{\em Phys. Rev. D} {\bfseries 87} (2013) 083008}, \href{https://arxiv.org/abs/1104.2333}{{\ttfamily arXiv:1104.2333 [astro-ph.CO]}}.

\bibitem{Baumann:2015rya}
D.~Baumann, D.~Green, J.~Meyers, and B.~Wallisch, ``{Phases of New Physics in the CMB},'' \href{https://dx.doi.org/10.1088/1475-7516/2016/01/007}{{\em JCAP} {\bfseries 01} (2016) 007}, \href{https://arxiv.org/abs/1508.06342}{{\ttfamily arXiv:1508.06342 [astro-ph.CO]}}.

\bibitem{Nollett:2013pwa}
K.~M. Nollett and G.~Steigman, ``{BBN And The CMB Constrain Light, Electromagnetically Coupled WIMPs},'' \href{https://dx.doi.org/10.1103/PhysRevD.89.083508}{{\em Phys. Rev. D} {\bfseries 89} no.~8, (2014) 083508}, \href{https://arxiv.org/abs/1312.5725}{{\ttfamily arXiv:1312.5725 [astro-ph.CO]}}.

\bibitem{Nollett:2014lwa}
K.~M. Nollett and G.~Steigman, ``{BBN And The CMB Constrain Neutrino Coupled Light WIMPs},'' \href{https://dx.doi.org/10.1103/PhysRevD.91.083505}{{\em Phys. Rev. D} {\bfseries 91} no.~8, (2015) 083505}, \href{https://arxiv.org/abs/1411.6005}{{\ttfamily arXiv:1411.6005 [astro-ph.CO]}}.

\bibitem{Giovanetti:2024eff}
C.~Giovanetti, M.~Lisanti, H.~Liu, S.~Mishra-Sharma, and J.~T. Ruderman, ``{Cosmological Parameter Estimation with a Joint-Likelihood Analysis of the Cosmic Microwave Background and Big Bang Nucleosynthesis},'' \href{https://arxiv.org/abs/2408.14531}{{\ttfamily arXiv:2408.14531 [astro-ph.CO]}}.

\bibitem{Sikivie:1983ip}
P.~Sikivie, ``{Experimental Tests of the Invisible Axion},'' \href{https://dx.doi.org/10.1103/PhysRevLett.51.1415}{{\em Phys. Rev. Lett.} {\bfseries 51} (1983) 1415--1417}. [Erratum: Phys.Rev.Lett. 52, 695 (1984)].

\bibitem{vanBibber:1988ge}
K.~van Bibber, P.~M. McIntyre, D.~E. Morris, and G.~G. Raffelt, ``{A Practical Laboratory Detector for Solar Axions},'' \href{https://dx.doi.org/10.1103/PhysRevD.39.2089}{{\em Phys. Rev. D} {\bfseries 39} (1989) 2089}.

\bibitem{Redondo:2008aa}
J.~Redondo, ``{Helioscope Bounds on Hidden Sector Photons},'' \href{https://dx.doi.org/10.1088/1475-7516/2008/07/008}{{\em JCAP} {\bfseries 07} (2008) 008}, \href{https://arxiv.org/abs/0801.1527}{{\ttfamily arXiv:0801.1527 [hep-ph]}}.

\bibitem{Berlin:2021kcm}
A.~Berlin and K.~Schutz, ``{Helioscope for gravitationally bound millicharged particles},'' \href{https://dx.doi.org/10.1103/PhysRevD.105.095012}{{\em Phys. Rev. D} {\bfseries 105} no.~9, (2022) 095012}, \href{https://arxiv.org/abs/2111.01796}{{\ttfamily arXiv:2111.01796 [hep-ph]}}.

\bibitem{Irastorza:2011gs}
I.~G. Irastorza {\em et~al.}, ``{Towards a new generation axion helioscope},'' \href{https://dx.doi.org/10.1088/1475-7516/2011/06/013}{{\em JCAP} {\bfseries 06} (2011) 013}, \href{https://arxiv.org/abs/1103.5334}{{\ttfamily arXiv:1103.5334 [hep-ex]}}.

\bibitem{CAST:2017uph}
{\bfseries CAST} Collaboration, V.~Anastassopoulos {\em et~al.}, ``{New CAST Limit on the Axion-Photon Interaction},'' \href{https://dx.doi.org/10.1038/nphys4109}{{\em Nature Phys.} {\bfseries 13} (2017) 584--590}, \href{https://arxiv.org/abs/1705.02290}{{\ttfamily arXiv:1705.02290 [hep-ex]}}.

\bibitem{Cui:2017ytb}
Y.~Cui, M.~Pospelov, and J.~Pradler, ``{Signatures of Dark Radiation in Neutrino and Dark Matter Detectors},'' \href{https://dx.doi.org/10.1103/PhysRevD.97.103004}{{\em Phys. Rev. D} {\bfseries 97} no.~10, (2018) 103004}, \href{https://arxiv.org/abs/1711.04531}{{\ttfamily arXiv:1711.04531 [hep-ph]}}.

\bibitem{Kuo:2021mtp}
J.-L. Kuo, M.~Pospelov, and J.~Pradler, ``{Terrestrial probes of electromagnetically interacting dark radiation},'' \href{https://dx.doi.org/10.1103/PhysRevD.103.115030}{{\em Phys. Rev. D} {\bfseries 103} no.~11, (2021) 115030}, \href{https://arxiv.org/abs/2102.08409}{{\ttfamily arXiv:2102.08409 [hep-ph]}}.

\bibitem{Dror:2021nyr}
J.~A. Dror, H.~Murayama, and N.~L. Rodd, ``{Cosmic axion background},'' \href{https://dx.doi.org/10.1103/PhysRevD.103.115004}{{\em Phys. Rev. D} {\bfseries 103} no.~11, (2021) 115004}, \href{https://arxiv.org/abs/2101.09287}{{\ttfamily arXiv:2101.09287 [hep-ph]}}. [Erratum: Phys.Rev.D 106, 119902 (2022)].

\bibitem{ADMX:2023rsk}
{\bfseries ADMX} Collaboration, T.~Nitta {\em et~al.}, ``{Search for a Dark-Matter-Induced Cosmic Axion Background with ADMX},'' \href{https://dx.doi.org/10.1103/PhysRevLett.131.101002}{{\em Phys. Rev. Lett.} {\bfseries 131} no.~10, (2023) 101002}, \href{https://arxiv.org/abs/2303.06282}{{\ttfamily arXiv:2303.06282 [hep-ex]}}.

\bibitem{Holdom:1985ag}
B.~Holdom, ``{Two U(1)'s and Epsilon Charge Shifts},'' \href{https://dx.doi.org/10.1016/0370-2693(86)91377-8}{{\em Phys. Lett. B} {\bfseries 166} (1986) 196--198}.

\bibitem{Berlin:2019uco}
A.~Berlin, R.~T. D'Agnolo, S.~A.~R. Ellis, P.~Schuster, and N.~Toro, ``{Directly Deflecting Particle Dark Matter},'' \href{https://dx.doi.org/10.1103/PhysRevLett.124.011801}{{\em Phys. Rev. Lett.} {\bfseries 124} no.~1, (2020) 011801}, \href{https://arxiv.org/abs/1908.06982}{{\ttfamily arXiv:1908.06982 [hep-ph]}}.

\bibitem{Graham:2014sha}
P.~W. Graham, J.~Mardon, S.~Rajendran, and Y.~Zhao, ``{Parametrically enhanced hidden photon search},'' \href{https://dx.doi.org/10.1103/PhysRevD.90.075017}{{\em Phys. Rev. D} {\bfseries 90} no.~7, (2014) 075017}, \href{https://arxiv.org/abs/1407.4806}{{\ttfamily arXiv:1407.4806 [hep-ph]}}.

\bibitem{Romanenko:2023irv}
A.~Romanenko {\em et~al.}, ``{Search for Dark Photons with Superconducting Radio Frequency Cavities},'' \href{https://dx.doi.org/10.1103/PhysRevLett.130.261801}{{\em Phys. Rev. Lett.} {\bfseries 130} no.~26, (2023) 261801}, \href{https://arxiv.org/abs/2301.11512}{{\ttfamily arXiv:2301.11512 [hep-ex]}}.

\bibitem{Romanenko:2014yaa}
A.~Romanenko, A.~Grassellino, A.~C. Crawford, D.~A. Sergatskov, and O.~Melnychuk, ``{Ultra-high quality factors in superconducting niobium cavities in ambient magnetic fields up to 190 mG},'' \href{https://dx.doi.org/10.1063/1.4903808}{{\em Appl. Phys. Lett.} {\bfseries 105} (2014) 234103}, \href{https://arxiv.org/abs/1410.7877}{{\ttfamily arXiv:1410.7877 [physics.acc-ph]}}.

\bibitem{Berlin:2023gvx}
A.~Berlin, R.~Tito~D'Agnolo, S.~A.~R. Ellis, and J.~I. Radkovski, ``{Signals of millicharged dark matter in light-shining-through-wall experiments},'' \href{https://dx.doi.org/10.1007/JHEP08(2023)017}{{\em JHEP} {\bfseries 08} (2023) 017}, \href{https://arxiv.org/abs/2305.05684}{{\ttfamily arXiv:2305.05684 [hep-ph]}}.

\bibitem{Berlin:2020pey}
A.~Berlin and A.~Hook, ``{Searching for Millicharged Particles with Superconducting Radio-Frequency Cavities},'' \href{https://dx.doi.org/10.1103/PhysRevD.102.035010}{{\em Phys. Rev. D} {\bfseries 102} no.~3, (2020) 035010}, \href{https://arxiv.org/abs/2001.02679}{{\ttfamily arXiv:2001.02679 [hep-ph]}}.

\bibitem{Vinyoles:2015khy}
N.~Vinyoles and H.~Vogel, ``{Minicharged Particles from the Sun: A Cutting-Edge Bound},'' \href{https://dx.doi.org/10.1088/1475-7516/2016/03/002}{{\em JCAP} {\bfseries 03} (2016) 002}, \href{https://arxiv.org/abs/1511.01122}{{\ttfamily arXiv:1511.01122 [hep-ph]}}.

\bibitem{Fung:2023euv}
A.~Fung, S.~Heeba, Q.~Liu, V.~Muralidharan, K.~Schutz, and A.~C. Vincent, ``{New bounds on light millicharged particles from the tip of the red-giant branch},'' \href{https://dx.doi.org/10.1103/PhysRevD.109.083011}{{\em Phys. Rev. D} {\bfseries 109} no.~8, (2024) 083011}, \href{https://arxiv.org/abs/2309.06465}{{\ttfamily arXiv:2309.06465 [hep-ph]}}.

\bibitem{Vernet:debye}
N.~Meyer‐Vernet, ``{Aspects of Debye shielding},'' \href{https://dx.doi.org/10.1119/1.17300}{{\em American Journal of Physics} {\bfseries 61} no.~3, (03, 1993) 249--257}, \href{https://arxiv.org/abs/https://pubs.aip.org/aapt/ajp/article-pdf/61/3/249/12179178/249\_1\_online.pdf}{{\ttfamily https://pubs.aip.org/aapt/ajp/article-pdf/61/3/249/12179178/249\_1\_online.pdf}}. \url{https://doi.org/10.1119/1.17300}.

\bibitem{Chang:2022gcs}
J.~H. Chang, D.~E. Kaplan, S.~Rajendran, H.~Ramani, and E.~H. Tanin, ``{Dark Solar Wind},'' \href{https://dx.doi.org/10.1103/PhysRevLett.129.211101}{{\em Phys. Rev. Lett.} {\bfseries 129} no.~21, (2022) 211101}, \href{https://arxiv.org/abs/2205.11527}{{\ttfamily arXiv:2205.11527 [hep-ph]}}.

\bibitem{Thoma:2008my}
M.~H. Thoma, ``{Field Theoretic Description of Ultrarelativistic Electron-Positron Plasmas},'' \href{https://dx.doi.org/10.1103/RevModPhys.81.959}{{\em Rev. Mod. Phys.} {\bfseries 81} (2009) 959--968}, \href{https://arxiv.org/abs/0801.0956}{{\ttfamily arXiv:0801.0956 [physics.plasm-ph]}}.

\bibitem{lifschitz1983physical}
E.~Lifschitz and L.~Pitajewski, ``Physical kinetics,'' in {\em Textbook of theoretical physics. 10}.
\newblock 1983.

\bibitem{Blaizot:2001nr}
J.-P. Blaizot and E.~Iancu, ``{The Quark gluon plasma: Collective dynamics and hard thermal loops},'' \href{https://dx.doi.org/10.1016/S0370-1573(01)00061-8}{{\em Phys. Rept.} {\bfseries 359} (2002) 355--528}, \href{https://arxiv.org/abs/hep-ph/0101103}{{\ttfamily arXiv:hep-ph/0101103}}.

\bibitem{Thoma:1994fd}
M.~H. Thoma, ``{Damping rate of a hard photon in a relativistic plasma},'' \href{https://dx.doi.org/10.1103/PhysRevD.51.862}{{\em Phys. Rev. D} {\bfseries 51} (1995) 862--865}, \href{https://arxiv.org/abs/hep-ph/9405309}{{\ttfamily arXiv:hep-ph/9405309}}.

\bibitem{Mrowczynski:1989np}
S.~Mrowczynski, ``{KINETIC THEORY APPROACH TO QUARK - GLUON PLASMA OSCILLATIONS},'' \href{https://dx.doi.org/10.1103/PhysRevD.39.1940}{{\em Phys. Rev. D} {\bfseries 39} (1989) 1940--1946}.

\bibitem{Baym:1997gq}
G.~Baym and H.~Heiselberg, ``{The Electrical conductivity in the early universe},'' \href{https://dx.doi.org/10.1103/PhysRevD.56.5254}{{\em Phys. Rev. D} {\bfseries 56} (1997) 5254--5259}, \href{https://arxiv.org/abs/astro-ph/9704214}{{\ttfamily arXiv:astro-ph/9704214}}.

\bibitem{Silin:1960pya}
V.~P. Silin, ``{On the electronmagnetic properties of a relativistic plasma},'' {\em Sov. Phys. JETP} {\bfseries 11} no.~5, (1960) 1136--1140.

\bibitem{Klimov:1982bv}
V.~V. Klimov, ``{Collective Excitations in a Hot Quark Gluon Plasma},'' {\em Sov. Phys. JETP} {\bfseries 55} (1982) 199--204.

\bibitem{Weldon:1982aq}
H.~A. Weldon, ``{Covariant Calculations at Finite Temperature: The Relativistic Plasma},'' \href{https://dx.doi.org/10.1103/PhysRevD.26.1394}{{\em Phys. Rev. D} {\bfseries 26} (1982) 1394}.

\bibitem{Kelly:1994dh}
P.~F. Kelly, Q.~Liu, C.~Lucchesi, and C.~Manuel, ``{Classical transport theory and hard thermal loops in the quark - gluon plasma},'' \href{https://dx.doi.org/10.1103/PhysRevD.50.4209}{{\em Phys. Rev. D} {\bfseries 50} (1994) 4209--4218}, \href{https://arxiv.org/abs/hep-ph/9406285}{{\ttfamily arXiv:hep-ph/9406285}}.

\bibitem{Kelly:1994ig}
P.~F. Kelly, Q.~Liu, C.~Lucchesi, and C.~Manuel, ``{Deriving the hard thermal loops of QCD from classical transport theory},'' \href{https://dx.doi.org/10.1103/PhysRevLett.72.3461}{{\em Phys. Rev. Lett.} {\bfseries 72} (1994) 3461--3463}, \href{https://arxiv.org/abs/hep-ph/9403403}{{\ttfamily arXiv:hep-ph/9403403}}.

\bibitem{Blaizot:1993zk}
J.~P. Blaizot and E.~Iancu, ``{Kinetic equations for long wavelength excitations of the quark - gluon plasma},'' \href{https://dx.doi.org/10.1103/PhysRevLett.70.3376}{{\em Phys. Rev. Lett.} {\bfseries 70} (1993) 3376--3379}, \href{https://arxiv.org/abs/hep-ph/9301236}{{\ttfamily arXiv:hep-ph/9301236}}.

\bibitem{melrose2008quantum}
D.~B. Melrose, {\em Quantum plasmadynamics: unmagnetized plasmas}, vol.~735.
\newblock Springer, 2008.

\bibitem{Chu:1988wh}
M.~C. Chu and T.~Matsui, ``{Dynamic Debye Screening for a Heavy Anti-quark Pair Traversing a Quark - Gluon Plasma},'' \href{https://dx.doi.org/10.1103/PhysRevD.39.1892}{{\em Phys. Rev. D} {\bfseries 39} (1989) 1892}.

\bibitem{Chakraborty:2006md}
P.~Chakraborty, M.~G. Mustafa, and M.~H. Thoma, ``{Wakes in the quark-gluon plasma},'' \href{https://dx.doi.org/10.1103/PhysRevD.74.094002}{{\em Phys. Rev. D} {\bfseries 74} (2006) 094002}, \href{https://arxiv.org/abs/hep-ph/0606316}{{\ttfamily arXiv:hep-ph/0606316}}.

\bibitem{Hill}
D.~Hill, {\em {Electromagnetic Fields in Cavities: Deterministic and Statistical Theories}}.
\newblock Wiley/IEEE Press, Piscataway, NJ, 2009-09-21, 2009.

\bibitem{Dennis:2023kfe}
M.~T. Dennis and J.~Sakstein, ``{Tip of the Red Giant Branch Bounds on the Axion-Electron Coupling Revisited},'' \href{https://arxiv.org/abs/2305.03113}{{\ttfamily arXiv:2305.03113 [hep-ph]}}.

\bibitem{Caputo:2024oqc}
A.~Caputo and G.~Raffelt, ``{Astrophysical Axion Bounds: The 2024 Edition},'' \href{https://dx.doi.org/10.22323/1.454.0041}{{\em PoS} {\bfseries COSMICWISPers} (2024) 041}, \href{https://arxiv.org/abs/2401.13728}{{\ttfamily arXiv:2401.13728 [hep-ph]}}.

\bibitem{Chang:2019xva}
J.~H. Chang, R.~Essig, and A.~Reinert, ``{Light(ly)-coupled Dark Matter in the keV Range: Freeze-In and Constraints},'' \href{https://dx.doi.org/10.1007/JHEP03(2021)141}{{\em JHEP} {\bfseries 03} (2021) 141}, \href{https://arxiv.org/abs/1911.03389}{{\ttfamily arXiv:1911.03389 [hep-ph]}}.

\bibitem{Planck:2018vyg}
{\bfseries Planck} Collaboration, N.~Aghanim {\em et~al.}, ``{Planck 2018 results. VI. Cosmological parameters},'' \href{https://dx.doi.org/10.1051/0004-6361/201833910}{{\em Astron. Astrophys.} {\bfseries 641} (2020) A6}, \href{https://arxiv.org/abs/1807.06209}{{\ttfamily arXiv:1807.06209 [astro-ph.CO]}}. [Erratum: Astron.Astrophys. 652, C4 (2021)].

\bibitem{ACT:2020gnv}
{\bfseries ACT} Collaboration, S.~Aiola {\em et~al.}, ``{The Atacama Cosmology Telescope: DR4 Maps and Cosmological Parameters},'' \href{https://dx.doi.org/10.1088/1475-7516/2020/12/047}{{\em JCAP} {\bfseries 12} (2020) 047}, \href{https://arxiv.org/abs/2007.07288}{{\ttfamily arXiv:2007.07288 [astro-ph.CO]}}.

\bibitem{Blinov:2020hmc}
N.~Blinov and G.~Marques-Tavares, ``{Interacting radiation after Planck and its implications for the Hubble Tension},'' \href{https://dx.doi.org/10.1088/1475-7516/2020/09/029}{{\em JCAP} {\bfseries 09} (2020) 029}, \href{https://arxiv.org/abs/2003.08387}{{\ttfamily arXiv:2003.08387 [astro-ph.CO]}}.

\bibitem{Dvorkin:2019zdi}
C.~Dvorkin, T.~Lin, and K.~Schutz, ``{Making dark matter out of light: freeze-in from plasma effects},'' \href{https://dx.doi.org/10.1103/PhysRevD.99.115009}{{\em Phys. Rev. D} {\bfseries 99} no.~11, (2019) 115009}, \href{https://arxiv.org/abs/1902.08623}{{\ttfamily arXiv:1902.08623 [hep-ph]}}. [Erratum: Phys.Rev.D 105, 119901 (2022)].

\bibitem{Berlin:2022hmt}
A.~Berlin, J.~A. Dror, X.~Gan, and J.~T. Ruderman, ``{Millicharged relics reveal massless dark photons},'' \href{https://dx.doi.org/10.1007/JHEP05(2023)046}{{\em JHEP} {\bfseries 05} (2023) 046}, \href{https://arxiv.org/abs/2211.05139}{{\ttfamily arXiv:2211.05139 [hep-ph]}}.

\bibitem{DES:2020mpv}
{\bfseries DES} Collaboration, A.~Chen {\em et~al.}, ``{Constraints on dark matter to dark radiation conversion in the late universe with DES-Y1 and external data},'' \href{https://dx.doi.org/10.1103/PhysRevD.103.123528}{{\em Phys. Rev. D} {\bfseries 103} no.~12, (2021) 123528}, \href{https://arxiv.org/abs/2011.04606}{{\ttfamily arXiv:2011.04606 [astro-ph.CO]}}.

\bibitem{Safdi:2022xkm}
B.~R. Safdi, ``{TASI Lectures on the Particle Physics and Astrophysics of Dark Matter},'' \href{https://dx.doi.org/10.22323/1.439.0009}{{\em PoS} {\bfseries TASI2022} (2024) 009}, \href{https://arxiv.org/abs/2303.02169}{{\ttfamily arXiv:2303.02169 [hep-ph]}}.

\bibitem{Berghaus:2020ekh}
K.~V. Berghaus, P.~W. Graham, D.~E. Kaplan, G.~D. Moore, and S.~Rajendran, ``{Dark energy radiation},'' \href{https://dx.doi.org/10.1103/PhysRevD.104.083520}{{\em Phys. Rev. D} {\bfseries 104} no.~8, (2021) 083520}, \href{https://arxiv.org/abs/2012.10549}{{\ttfamily arXiv:2012.10549 [hep-ph]}}.

\bibitem{Ji:2021mvg}
L.~Ji, D.~E. Kaplan, S.~Rajendran, and E.~H. Tanin, ``{Thermal perturbations from cosmological constant relaxation},'' \href{https://dx.doi.org/10.1103/PhysRevD.105.015025}{{\em Phys. Rev. D} {\bfseries 105} no.~1, (2022) 015025}, \href{https://arxiv.org/abs/2109.05285}{{\ttfamily arXiv:2109.05285 [hep-ph]}}.

\bibitem{Berghaus:2023ypi}
K.~V. Berghaus, T.~Karwal, V.~Miranda, and T.~Brinckmann, ``{The Cosmology of Dark Energy Radiation},'' \href{https://arxiv.org/abs/2311.08638}{{\ttfamily arXiv:2311.08638 [hep-ph]}}.

\bibitem{Fox:2011qd}
P.~J. Fox, J.~Liu, D.~Tucker-Smith, and N.~Weiner, ``{An Effective Z'},'' \href{https://dx.doi.org/10.1103/PhysRevD.84.115006}{{\em Phys. Rev. D} {\bfseries 84} (2011) 115006}, \href{https://arxiv.org/abs/1104.4127}{{\ttfamily arXiv:1104.4127 [hep-ph]}}.

\bibitem{Schmaltz:2017oov}
M.~Schmaltz and N.~Weiner, ``{A Portalino to the Dark Sector},'' \href{https://dx.doi.org/10.1007/JHEP02(2019)105}{{\em JHEP} {\bfseries 02} (2019) 105}, \href{https://arxiv.org/abs/1709.09164}{{\ttfamily arXiv:1709.09164 [hep-ph]}}.

\bibitem{Giunti:2014ixa}
C.~Giunti and A.~Studenikin, ``{Neutrino electromagnetic interactions: a window to new physics},'' \href{https://dx.doi.org/10.1103/RevModPhys.87.531}{{\em Rev. Mod. Phys.} {\bfseries 87} (2015) 531}, \href{https://arxiv.org/abs/1403.6344}{{\ttfamily arXiv:1403.6344 [hep-ph]}}.

\bibitem{Giunti:2015gga}
C.~Giunti, K.~A. Kouzakov, Y.-F. Li, A.~V. Lokhov, A.~I. Studenikin, and S.~Zhou, ``{Electromagnetic neutrinos in laboratory experiments and astrophysics},'' \href{https://dx.doi.org/10.1002/andp.201500211}{{\em Annalen Phys.} {\bfseries 528} (2016) 198--215}, \href{https://arxiv.org/abs/1506.05387}{{\ttfamily arXiv:1506.05387 [hep-ph]}}.

\bibitem{Giunti:2024gec}
C.~Giunti, K.~Kouzakov, Y.-F. Li, and A.~Studenikin, ``{Neutrino Electromagnetic Properties},'' \href{https://arxiv.org/abs/2411.03122}{{\ttfamily arXiv:2411.03122 [hep-ph]}}.

\bibitem{Raffelt:1999gv}
G.~G. Raffelt, ``{Limits on neutrino electromagnetic properties: An update},'' \href{https://dx.doi.org/10.1016/S0370-1573(99)00074-5}{{\em Phys. Rept.} {\bfseries 320} (1999) 319--327}.

\bibitem{Barbiellini:1987zz}
G.~Barbiellini and G.~Cocconi, ``{Electric Charge of the Neutrinos from SN1987A},'' \href{https://dx.doi.org/10.1038/329021b0}{{\em Nature} {\bfseries 329} (1987) 21--22}.

\bibitem{Green:2017ybv}
D.~Green and S.~Rajendran, ``{The Cosmology of Sub-MeV Dark Matter},'' \href{https://dx.doi.org/10.1007/JHEP10(2017)013}{{\em JHEP} {\bfseries 10} (2017) 013}, \href{https://arxiv.org/abs/1701.08750}{{\ttfamily arXiv:1701.08750 [hep-ph]}}.

\bibitem{Berlin:2017ftj}
A.~Berlin and N.~Blinov, ``{Thermal Dark Matter Below an MeV},'' \href{https://dx.doi.org/10.1103/PhysRevLett.120.021801}{{\em Phys. Rev. Lett.} {\bfseries 120} no.~2, (2018) 021801}, \href{https://arxiv.org/abs/1706.07046}{{\ttfamily arXiv:1706.07046 [hep-ph]}}.

\bibitem{Berlin:2018ztp}
A.~Berlin and N.~Blinov, ``{Thermal neutrino portal to sub-MeV dark matter},'' \href{https://dx.doi.org/10.1103/PhysRevD.99.095030}{{\em Phys. Rev. D} {\bfseries 99} no.~9, (2019) 095030}, \href{https://arxiv.org/abs/1807.04282}{{\ttfamily arXiv:1807.04282 [hep-ph]}}.

\bibitem{Beacom:2004yd}
J.~F. Beacom, N.~F. Bell, and S.~Dodelson, ``{Neutrinoless universe},'' \href{https://dx.doi.org/10.1103/PhysRevLett.93.121302}{{\em Phys. Rev. Lett.} {\bfseries 93} (2004) 121302}, \href{https://arxiv.org/abs/astro-ph/0404585}{{\ttfamily arXiv:astro-ph/0404585}}.

\bibitem{Emken:2019tni}
T.~Emken, R.~Essig, C.~Kouvaris, and M.~Sholapurkar, ``{Direct Detection of Strongly Interacting Sub-GeV Dark Matter via Electron Recoils},'' \href{https://dx.doi.org/10.1088/1475-7516/2019/09/070}{{\em JCAP} {\bfseries 09} (2019) 070}, \href{https://arxiv.org/abs/1905.06348}{{\ttfamily arXiv:1905.06348 [hep-ph]}}.

\bibitem{Cholis:2015gna}
I.~Cholis, D.~Hooper, and T.~Linden, ``{A Predictive Analytic Model for the Solar Modulation of Cosmic Rays},'' \href{https://dx.doi.org/10.1103/PhysRevD.93.043016}{{\em Phys. Rev. D} {\bfseries 93} no.~4, (2016) 043016}, \href{https://arxiv.org/abs/1511.01507}{{\ttfamily arXiv:1511.01507 [astro-ph.SR]}}.

\bibitem{Potgieter:2013mcc}
M.~S. Potgieter, ``{Very local interstellar spectra for galactic electrons, protons and helium},'' \href{https://dx.doi.org/10.1007/s13538-014-0238-2}{{\em Braz. J. Phys.} {\bfseries 44} (2014) 581--588}, \href{https://arxiv.org/abs/1310.6133}{{\ttfamily arXiv:1310.6133 [astro-ph.SR]}}.

\bibitem{Chuzhoy:2008zy}
L.~Chuzhoy and E.~W. Kolb, ``{Reopening the window on charged dark matter},'' \href{https://dx.doi.org/10.1088/1475-7516/2009/07/014}{{\em JCAP} {\bfseries 07} (2009) 014}, \href{https://arxiv.org/abs/0809.0436}{{\ttfamily arXiv:0809.0436 [astro-ph]}}.

\bibitem{Li:2020wyl}
J.-T. Li and T.~Lin, ``{Dynamics of millicharged dark matter in supernova remnants},'' \href{https://dx.doi.org/10.1103/PhysRevD.101.103034}{{\em Phys. Rev. D} {\bfseries 101} no.~10, (2020) 103034}, \href{https://arxiv.org/abs/2002.04625}{{\ttfamily arXiv:2002.04625 [astro-ph.CO]}}.

\bibitem{Ruppert:2005uz}
J.~Ruppert and B.~Muller, ``{Waking the colored plasma},'' \href{https://dx.doi.org/10.1016/j.physletb.2005.04.075}{{\em Phys. Lett. B} {\bfseries 618} (2005) 123--130}, \href{https://arxiv.org/abs/hep-ph/0503158}{{\ttfamily arXiv:hep-ph/0503158}}.

\bibitem{Thoma:2008tb}
M.~H. Thoma, ``{Ultrarelativistic Electron-Positron Plasma},'' \href{https://dx.doi.org/10.1140/epjd/e2009-00077-9}{{\em Eur. Phys. J. D} {\bfseries 55} (2009) 271--278}, \href{https://arxiv.org/abs/0810.0909}{{\ttfamily arXiv:0810.0909 [hep-ph]}}.

\bibitem{Mrowczynski:2007hb}
S.~Mrowczynski and M.~H. Thoma, ``{What Do Electromagnetic Plasmas Tell Us about Quark-Gluon Plasma?},'' \href{https://dx.doi.org/10.1146/annurev.nucl.57.090506.123124}{{\em Ann. Rev. Nucl. Part. Sci.} {\bfseries 57} (2007) 61--94}, \href{https://arxiv.org/abs/nucl-th/0701002}{{\ttfamily arXiv:nucl-th/0701002}}.

\bibitem{Carrington:2003je}
M.~E. Carrington, T.~Fugleberg, D.~Pickering, and M.~H. Thoma, ``{Dielectric functions and dispersion relations of ultrarelativistic plasmas with collisions},'' \href{https://dx.doi.org/10.1139/p04-035}{{\em Can. J. Phys.} {\bfseries 82} (2004) 671--678}, \href{https://arxiv.org/abs/hep-ph/0312103}{{\ttfamily arXiv:hep-ph/0312103}}.

\bibitem{Mandal:2012wi}
M.~Mandal and P.~Roy, ``{Wake in anisotropic quark-gluon plasma},'' \href{https://dx.doi.org/10.1103/PhysRevD.86.114002}{{\em Phys. Rev. D} {\bfseries 86} (2012) 114002}, \href{https://arxiv.org/abs/1310.4657}{{\ttfamily arXiv:1310.4657 [hep-ph]}}.

\bibitem{Grayson:2022asf}
C.~Grayson, M.~Formanek, J.~Rafelski, and B.~Mueller, ``{Dynamic magnetic response of the quark-gluon plasma to electromagnetic fields},'' \href{https://dx.doi.org/10.1103/PhysRevD.106.014011}{{\em Phys. Rev. D} {\bfseries 106} no.~1, (2022) 014011}, \href{https://arxiv.org/abs/2204.14186}{{\ttfamily arXiv:2204.14186 [hep-ph]}}.

\bibitem{Ahonen:1998iz}
J.~Ahonen, ``{Transport coefficients in the early universe},'' \href{https://dx.doi.org/10.1103/PhysRevD.59.023004}{{\em Phys. Rev. D} {\bfseries 59} (1999) 023004}, \href{https://arxiv.org/abs/hep-ph/9801434}{{\ttfamily arXiv:hep-ph/9801434}}.

\end{thebibliography}\endgroup

\end{document}